\documentclass[lettersize,journal]{IEEEtran}
\usepackage{amsmath,amsfonts}
\usepackage{array}
\usepackage[caption=false,font=normalsize,labelfont=sf,textfont=sf]{subfig}
\usepackage{textcomp}
\usepackage{stfloats}
\usepackage{url}
\usepackage{verbatim}
\usepackage{graphicx}
\usepackage{cite}
\usepackage{xcolor}
\usepackage[final,commandnameprefix=always]{changes}
\setauthormarkup{}
\definechangesauthor[color=blue]{osw}

\usepackage{tikz}
\usepackage{amsmath}

\usepackage{enumitem}
\usepackage{cite}

\usepackage[hidelinks]{hyperref}

\usepackage{listings}

\definecolor{codegreen}{rgb}{0,0.6,0}
\definecolor{codegray}{rgb}{0.5,0.5,0.5}
\definecolor{codepurple}{rgb}{0.58,0,0.82}
\definecolor{backcolour}{rgb}{0.95,0.95,0.92}

\lstdefinestyle{mystyle}{
    backgroundcolor=\color{backcolour},   
    commentstyle=\color{codegreen},
    keywordstyle=\color{magenta},
    numberstyle=\tiny\color{codegray},
    stringstyle=\color{codepurple},
    basicstyle=\ttfamily\footnotesize,
    breakatwhitespace=false,         
    breaklines=true,                 
    captionpos=b,                    
    keepspaces=true,                 
    numbers=left,                    
    numbersep=5pt,                  
    showspaces=false,                
    showstringspaces=false,
    showtabs=false,                  
    tabsize=2
}

\usepackage{xspace}
\newcommand{\sys}{{MirrorFuzz}\xspace}
\usepackage{booktabs}
\usepackage{tabularx}
\usepackage{adjustbox}
\usepackage{multirow}
\usepackage[linesnumbered, ruled, vlined]{algorithm2e}
\SetKwRepeat{Do}{do}{while}%

\begin{document}

\title{\sys: Leveraging LLM and Shared Bugs for Deep Learning Framework APIs Fuzzing}

\author{Shiwen Ou,
        Yuwei Li,
        Lu Yu,
        Chengkun Wei,
        Tingke Wen,
        Qiangpu Chen,\\
        Yu Chen,
        Haizhi Tang,
        and Zulie Pan
\thanks{Shiwen Ou, Yuwei Li, Lu Yu, Tingke Wen, Qiangpu Chen, Yu Chen, Haizhi Tang, and Zulie Pan are with the College of Electronic Engineering, National University of Defense Technology, Hefei 230037, China (e-mail: \{oushiwen, liyuwei, yulu, wentk, cy, panzulie17\}@nudt.edu.cn; \{chenqiangpu, tanghaizhi\}@foxmail.com).}
\thanks{Chengkun Wei is with the College of Computer Science and Technology, Zhejiang University, Hangzhou 310027, China (e-mail: weichengkun@zju.edu.cn).}}

\markboth{Journal of \LaTeX\ Class Files,~Vol.~14, No.~8, August~2021}%
{Shell \MakeLowercase{\textit{et al.}}: A Sample Article Using IEEEtran.cls for IEEE Journals}

\IEEEpubid{0000--0000/00\$00.00~\copyright~2021 IEEE}

\maketitle

\begin{abstract}
Deep learning (DL) frameworks serve as the backbone for a wide range of artificial intelligence applications.
However, bugs within DL frameworks can cascade into critical issues in higher-level applications, jeopardizing reliability and security.
While numerous techniques have been proposed to detect bugs in DL frameworks, research exploring common API patterns across frameworks and the potential risks they entail remains limited.
Notably, many DL frameworks expose similar APIs with overlapping input parameters and functionalities, rendering them vulnerable to shared bugs, where a flaw in one API may extend to analogous APIs in other frameworks.
To address this challenge,  we propose \sys, an automated API fuzzing solution to discover shared bugs in DL frameworks. 
\sys operates in three stages: First, \sys collects historical bug data for each API within a DL framework to identify potentially buggy APIs.
Second, it matches each buggy API in a specific framework with similar APIs within and across other DL frameworks.
Third, it employs large language models (LLMs) to synthesize code for the API under test, leveraging the historical bug data of similar APIs to trigger analogous bugs across APIs.
We implement \sys and evaluate it on four popular DL frameworks (TensorFlow, PyTorch, OneFlow, and Jittor).
Extensive evaluation demonstrates that \sys improves code coverage by 39.92\% and 98.20\% compared to state-of-the-art methods on TensorFlow and PyTorch, respectively.
Moreover, \sys discovers 315 bugs, 262 of which are newly found, and 80 bugs are fixed, with 52 of these bugs assigned CNVD IDs.
\end{abstract}

\section{Introduction}
\IEEEPARstart{D}{eep} learning techniques significantly impact real-world scenarios, including healthcare~\cite{shamshirband2021review,cascella2023evaluating}, education~\cite{kazemitabaar2024codeaid,balse2023evaluating}, finance~\cite{li2023large}, and autonomous driving~\cite{grigorescu2020survey}. Recent studies~\cite{go2024towards,christou2023ivysyn,deng2023differential,guan2023comprehensive,tambon2024silent, zhu2024my} highlight that DL frameworks (e.g., PyTorch~\cite{paszke2019PyTorch} and TensorFlow~\cite{abadi2016tensorflow}), which serve as the foundational infrastructure for various DL applications, are vulnerable to security flaws. These flaws can result in unexpected consequences in higher-layer applications, such as performance degradation~\cite{deng2023differential, humbatova2020taxonomy}, incorrect functionality~\cite{islam2019comprehensive, chen2023toward}, and memory errors~\cite{christou2023ivysyn}.

\subsection{Shared Bugs: Threats from Commonalities between DL Frameworks}
\label{subsection:Threats_from_Cross}
There is a surge of research on testing DL frameworks~\cite{zhang2024survey}, which predominantly focuses on two categories: model-level and API-level fuzzing.
Model-level fuzzing~\cite{pham2019cradle, wang2020deep, guo2020audee, gu2022muffin, liu2023generation} builds DL models as test cases and uses differential testing to identify defects, but it can only cover a limited set of APIs.
In contrast, API-level fuzzing~\cite{wei2022free, deng2022fuzzingDeepREL, xie2022docter, wang2022eagle, christou2023ivysyn, yang2023fuzzing, shi2023acetest, deng2023differential, deng2023large, deng2024largefuzzgpt, shirihistory} generates valid API invocation sequences, enabling broader API coverage without the need to construct full DL models, thus attracting attention.
Recent advancements in Large Language Models (LLMs) have also introduced LLM-based techniques to enhance fuzz testing of DL frameworks and beyond \cite{deng2023large, deng2024largefuzzgpt, go2024towards, guan2024large, oliinyk2024fuzzing, li2024seeds}. 
However, while these techniques have proven to be effective, they overlook an important factor: the commonalities between different DL frameworks. The security risks associated with similar APIs within and across frameworks remain under-explored.

\begin{figure}[t]
    \centering
    \includegraphics[width=1\linewidth]{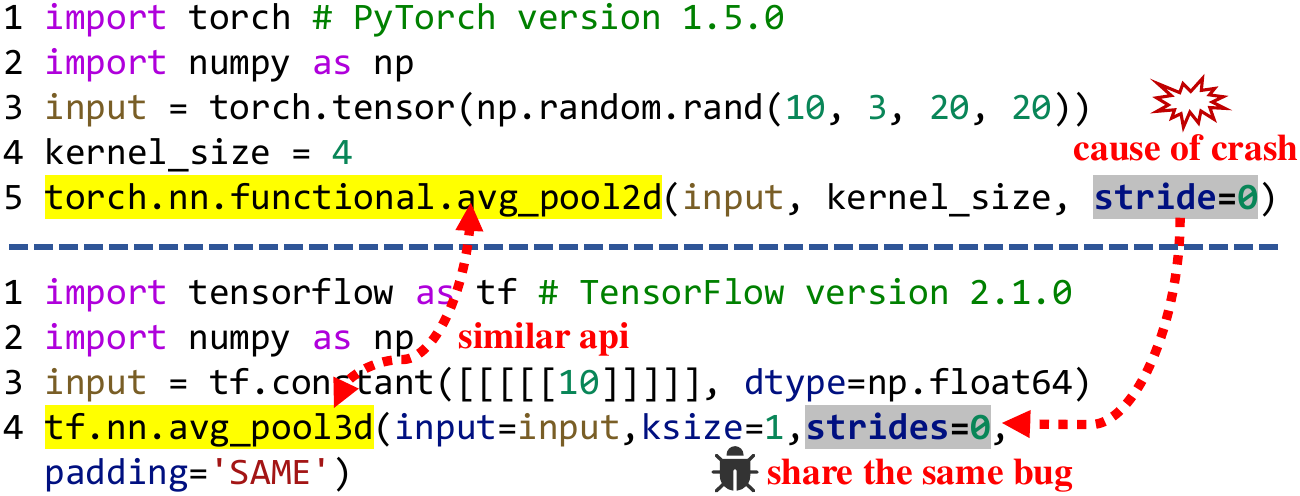}
    \caption{Examples of shared bugs.}
    \label{fig:examples_shared_bugs}
\end{figure}

\IEEEpubidadjcol \textit{Observations: Shared Bugs Within or Across DL frameworks.} It is worth noting that vulnerabilities in one software system can often be identified by referencing known issues in other systems that share or reuse source code, libraries, or specifications~\cite{pham2010detection, kang2022tracer}.
With the hypothesis that sharing software vulnerabilities, referred to as \textbf{shared bugs}, we conducted an empirical study on several DL frameworks, including PyTorch and TensorFlow.
We observe that DL frameworks exhibit similarities in features, architecture, and design, with comparable APIs prone to analogous bugs.
The shared bugs can enable early detection within or across DL framework APIs.
For instance, as shown in Figure~\ref{fig:examples_shared_bugs}, PyTorch's \texttt{torch.nn.functional.avg\_pool2d} and TensorFlow's \texttt{tf.nn.avg\_pool3d} APIs for pooling operations are susceptible to the same issue.
Setting the \texttt{stride} parameter in \texttt{avg\_pool2d} to 0, an invalid value, causes a crash in PyTorch, which can similarly be reproduced in TensorFlow's \texttt{avg\_pool3d} API under equivalent conditions.

However, existing DL framework fuzzing approaches overlook the similarities between DL frameworks and the potential security risks these similarities may introduce and fail to leverage shared bugs for early bug detection.

\noindent \textbf{Challenges.}
In this work, we aim to fuzz the APIs of DL frameworks by leveraging shared bugs.
To achieve this, we have to solve the following challenges.

\textit{C1: How to recognize the buggy API in the bug report?} The reported issues repository (e.g., GitHub) contains extensive bug reports related to framework APIs. However, recognizing buggy APIs is challenging due to deviations from community standards, ambiguous descriptions, incomplete code, and inconsistent report formats across different DL frameworks.

\textit{C2: How to match similar APIs within and across frameworks?}
Due to the diversity of APIs (such as differences in documentation, naming conventions, parameters, and return types), it is difficult to match similar APIs within and across frameworks. Even when two APIs are functionally equivalent, their details may still differ across frameworks.

\textit{C3: How to leverage shared bugs to synthesize test case that triggers similar bugs?}
Bugs found in the APIs of one framework may also manifest in similar APIs of other frameworks. 
Nevertheless, crafting test cases (i.e., code with API calls) that adhere to the target framework’s API constraints while effectively reproducing such bugs remains a significant challenge.

\subsection{\sys: A Fuzzer that Leveraging Shared Bugs Across DL Framework APIs}

\noindent \textbf{Our Approach.}
To address these challenges, we introduce \sys, an automated fuzzing approach powered by LLM to detect bugs in DL frameworks.
We propose and implement two techniques for effectively uncovering shared bugs in DL framework APIs.
First, we leverage the strengths of LLM in fuzzing DL framework APIs. 
Specifically, we use LLM for semantic recognition of bug reports, extracting buggy APIs from reported issues.
In addition, the ability of the LLM to generate structured high-quality outputs that comply with the target's input constraints enables the creation of precise test cases.
Second, we exploit the characteristics of shared bugs to identify the threat that an API may be susceptible to bugs in similar APIs within and across frameworks.
This approach is based on the intuition that an input causing a crash in one software is likely to trigger similar failures in other software exhibiting comparable behavior or design patterns.
This allows us to efficiently determine if the same bug exists across multiple DL frameworks, saving significant time by reducing the need for extensive fuzzing.

We design \sys with three modules:  \textit{Buggy APIs Recognizer}, \textit{Similar APIs Matcher}, and \textit{Test case Synthesizer}.
The \textit{Recognizer} collects bug reports from GitHub and employs LLM to identify buggy APIs  (see Section \ref{subsection:recognizer}).
Next, the \textit{Matcher} identifies similar APIs across frameworks based on text and semantic similarity (see Section \ref{subsection:matcher}).
Notably, we classify the matched similar APIs into \textit{parameter-similar} APIs and \textit{operation-similar} APIs, according to shared bug manifestations (see Section \ref{subsection:Motivation}).
Finally, the \textit{Synthesizer} generates test cases to detect bugs by utilizing the outputs from the previous modules (see Section \ref{subsection:synthesizer}). 

\chadded[id=osw]{\noindent \textbf{Evaluation.} We implement a prototype of \sys and evaluate it on four popular DL frameworks, i.e., TensorFlow \cite{abadi2016tensorflow}, PyTorch \cite{paszke2019PyTorch}, OneFlow \cite{OneFlowgithub}, and Jittor \cite{hu2020jittor}.
Although these frameworks are extensively tested in prior studies, \sys still successfully discovered 315 bugs in these DL frameworks, of which 262 were previously unknown.
We responsibly report these bugs to the corresponding developers in time.
As a result, 52 bugs are assigned CNVD IDs, 180 bugs are officially confirmed, and 80 bugs are fixed by the developers.
We compare \sys with the state-of-the-art DL fuzzing methods, including TENSORSCOPE \cite{deng2023differential}, TitanFuzz \cite{deng2023large}, Orion \cite{shirihistory} and MoCo \cite{ji2024moco}.
MirrorFuzz improves code coverage by 39.92\% on TensorFlow and 98.20\% on PyTorch, compared to the best-performing baseline.
Under the same time budget, MirrorFuzz tests more APIs and detects more unique bugs than all baselines, demonstrating its stronger efficiency and bug-finding capability.
More importantly, our research shows that bugs in leading frameworks like TensorFlow and PyTorch can spread to many other frameworks.
Also, bugs in smaller or newer frameworks like OneFlow and Jittor may affect mainstream ones, revealing the broad impact of shared bugs across DL frameworks. In summary, we make the following contributions:}
\begin{enumerate}
    \item We reveal shared bugs within and across DL frameworks.
    The high functional similarity across frameworks leads to many similar APIs.
    More importantly, APIs within a framework are also vulnerable to bugs found in other similar APIs, i.e., similar bugs may be shared across frameworks.
    \item We implement \sys, an automated fuzzing solution that leverages shared bugs and LLM to detect flaws in similar APIs within and across frameworks.
    Using LLMs' understanding of DL frameworks, we identify buggy APIs from bug reports and generate test cases to expose analogous bugs in target APIs.
    We open-source the code of \sys, the bug list, and data in \cite{mirrorfuzz_github}.
    \item We evaluate \sys on four popular DL frameworks and discover a total of 315 bugs.
    Among these, 262 are newly discovered, and 180 are officially confirmed.
    We also obtain 52 CNVD IDs.
\end{enumerate}

\section{Background \& Problem Statement}

\subsection{Fuzzing DL Framework}
\noindent \textbf{Basics about DL Framework.}
DL frameworks are designed to construct various DL models and applications.
Specifically, these frameworks provide essential building blocks for designing, training, and validating DL models through a wide variety of APIs. 
Thus, developers create DL programs using the provided APIs, which can then be executed with training data to build models and develop applications~\cite{chen2023toward}. 
With features such as automatic differentiation, GPU/TPU hardware acceleration, and efficient data manipulation, DL frameworks significantly enhance the efficiency and flexibility of developing and deploying DL applications.
With the growing diversity of requirements, numerous open-source DL frameworks~\cite{lenton2021ivy, Candlerust, flax2020github} have emerged.
These frameworks commonly provide core functionalities such as tensor computation, automatic differentiation, and model training.

\noindent \textbf{Model-level and API-level Fuzzing.}
Bug detection in DL frameworks mainly employs fuzzing techniques, which are divided into model-level and API-level approaches.
The model-level fuzzing approach primarily involves generating arbitrary complete DL models and feeding them into different backends, such as various frameworks (e.g., TensorFlow and PyTorch) or computing devices (e.g., CPU/GPU), using differential testing to detect crashes, inconsistencies, or performance issues within the frameworks.
However, its reliance on DL models as input in differential testing limits API coverage and is affected by precision loss~\cite{deng2022fuzzingDeepREL}.
In contrast, the API-level approach focuses on a specific API and generates complex API sequences to discover bugs within the framework by satisfying the constraints inferred from the API.
The API-level fuzzing approach removes the need to build complete DL models, enabling broader coverage of framework APIs, and has thus garnered widespread attention \cite{zhang2024survey}.
\begin{lstlisting}[language=Python, caption=Example of shared bugs in operation-similar APIs across DL frameworks., label=sim_op_diff, float=t,emph={output_size}, emphstyle={\bfseries\color{red}}]
# A bug in the torch.nn.AdaptiveMaxPool2d API.
import torch
size = 2 ** 32
m = torch.nn.AdaptiveMaxPool2d(output_size=size, return_indices=False)
inputs = torch.randn(1, 64, 8, 9)
m(inputs)  # bug

# A bug in the AdaptiveAvgPool1d API.
import oneflow as flow
size = 2 ** 32
m = flow.nn.AdaptiveAvgPool1d(output_size=size)
inputs = flow.Tensor(flow.randn(1, 64, 8))
m(inputs)  # bug
\end{lstlisting}
\begin{lstlisting}[language=Python, caption=Example of shared bugs in operation-similar APIs within the DL framework.,label=sim_op_sim, float=t,emph={input}, emphstyle={\bfseries\color{red}}]
# A bug in the torch.fft.irfftn API.
import torch
input = torch.randn(10, 512, 512)
output = torch.fft.irfftn(input)  # bug

# A bug in the torch.fft.hfftn API.
import torch
input = torch.randn(10, 512, 512)
output = torch.fft.hfftn(input)  # bug
\end{lstlisting}

\noindent \textbf{LLM for Fuzzing.}
As LLMs continue to evolve, researchers are using them for fuzz testing DL frameworks
\cite{deng2023large,deng2024largefuzzgpt,go2024towards,guan2024large}, for tasks such as seed generation, test case mutation, and constraint extraction.
LLMs trained on vast amounts of text data have demonstrated powerful performance in natural language processing tasks.
Additionally, some variants of LLMs (such as CodeLlama~\cite{roziere2023code},  CodeGemma~\cite{team2024codegemma}, and Qwen2.5-Coder~\cite{hui2024qwen2}) are also trained and fine-tuned on specific code snippet datasets, enabling them to handle code-related tasks (such as program analysis, code generation, etc.). 
This allows LLMs to generate code snippets that satisfy the complex syntax and semantic requirements of DL framework APIs \cite{deng2023large}.
\subsection{Problem Statement}
\label{subsection:Motivation}

\noindent \textbf{New Threat.}
DL framework APIs enable users to interact with the framework for tasks like dataset processing, model building, training, evaluation, and inference.
Interestingly, APIs from different frameworks are often developed independently but share similar functionality and design.
We observe significant commonalities across different DL framework APIs, driven by three main factors.
First, many DL frameworks share essential operations such as tensor computation, automatic differentiation, and dataset handling.
Second, the underlying logic of supported algorithms is consistent, such as the mathematical principles behind convolution and pooling in convolutional neural networks (CNNs\cite{lecun1998gradientcnn}). 
Finally, the enhancement of interoperability between frameworks has driven API consistency. For example, most frameworks now support the ONNX standard \cite{ONNXst} to facilitate model import, export, and conversion.
Due to code reuse and compatibility, there are also many similar APIs within the framework itself.
Inspired by the well-known hypothesis that similar systems may exhibit similar bugs, we speculate that APIs within and across DL frameworks may share bugs, and we investigate this through empirical study.

\noindent \textbf{Motivation Eaxmple.}
\label{subsection:sharedbug_example}
We collect API bug reports from GitHub Issues for four popular DL frameworks, including TensorFlow\cite{abadi2016tensorflow}, PyTorch\cite{paszke2019PyTorch}, OneFlow\cite{OneFlowgithub}, and Jittor\cite{hu2020jittor}.
We observe that a shared bug refers to a defect occurring in similar APIs within or across frameworks. These defects are often caused by common designs, logic, or implementation patterns, resulting in similar issues.
Moreover, we analyze the occurrence of these shared bugs as follows.

\noindent \textbf{Operation Similarity APIs.}
We find that APIs with similar operations often share similar bugs, both within and across frameworks.
For example, as shown in Listing \ref{sim_op_diff}, \texttt{torch.nn.AdaptiveMaxPool2d} and \texttt{flow.nn.AdaptiveAvgPool1d} triggered the same crash due to an excessively large \textit{output\_size} parameter.
While one performs max pooling and the other average pooling, both are pooling operations for downsampling feature maps.
Therefore, we categorize this type of API as \textit{operation-similar} APIs.
A pair of APIs is classified as \textit{operation-similar} if they perform similar functions and produce comparable outputs when provided with analogous inputs.
Similarly, Listing \ref{sim_op_sim} shows that \textit{operation-similar} APIs within the same framework can also share bugs.

\noindent \textbf{Parameter Similarity APIs.}
We observe that APIs within or across frameworks may share certain parameters, even though their functionality or operations differ significantly.
Interestingly, these shared parameters often result in similar bugs.
For example, both \texttt{torch.nn.Conv2d} and \texttt{torch.nn.functional.avg\_pool2d} crash when the \textit{stride} is set to zero.
Despite performing different operations (padding and pooling), the crash is triggered by the same parameter with the same underlying cause.
In PyTorch version 1.2, the bug in \texttt{torch.nn.Conv2d} gets fixed, and calling it with a zero stride raises a runtime error stating that ``non-positive stride is not supported''.
However, in version 1.5, \texttt{torch.nn.functional.avg\_pool2d} still crashes with a zero stride.
We further observe that even in the latest versions, APIs with stride parameters, such as \texttt{torch.ao.nn.quantized.Conv1d}, still crashes due to the same issue.
This highlights two points: 1) APIs with different functionalities but similar parameters may also share bugs. 2) fixing a bug in one API does not always address similar issues in others with the same parameter, as each API may implement its functionality independently.
Moreover, as shown in Listing \ref{para_diff}, this case also shares bugs between frameworks.
We classify such pairs of APIs as \textit{parameter-similar} APIs, which share comparable input parameters, even though they may perform different operations or tasks.

The \textit{parameter-similar} APIs serve as an important complement to \textit{operation-similar} APIs, and we provide definitions for both types of APIs in Section \ref{subsection:matcher}, along with the details of matching these two types of APIs.
\begin{lstlisting}[language=Python, caption=Example of shared bugs in parameter-similar APIs within the DL framework., label=para_sim, morekeywords={stride}, float=t,emph={stride}, emphstyle={\bfseries\color{red}}]
# A bug in the torch.nn.Conv2d API.
import torch
from torch import nn
input = torch.randn(1, 1, 32, 32)
c = nn.Conv2d(in_channels=1, out_channels=32, kernel_size=4, stride=0, bias=False, padding=(0, 1), padding_mode='constant')
c(input) # bug

# A bug in the avg_pool2d API.
import torch
input = torch.randn(10, 3, 20, 20)
torch.nn.functional.avg_pool2d(input, kernel_size=4, stride=0)  # bug
\end{lstlisting}
\begin{lstlisting}[language=Python, caption=Example of shared bugs in parameter-similar APIs across DL frameworks., label=para_diff, morekeywords={padding_num}, float=t, emph={padding}, emphstyle={\bfseries\color{red}}]
# A bug in the torch.nn.ReflectionPad2d API.
import torch
padding_num = -8353862602220610428
# bug
torch.nn.ReflectionPad2d(padding=padding_num)

# A bug in the flow.nn.MaxPool1d API. 
import oneflow as flow
padding_num= -8353862602220610428
pool = flow.nn.MaxPool1d(2, stride=2, return_indices=True,padding=padding_num)
input = flow.tensor([[[1., 2, 3, 4, 5, 6, 7, 8]]])
output, indices = pool(input) # bug
\end{lstlisting}
\section{Approach}
\begin{figure}[t]
    \centering
    \includegraphics[width=1\linewidth]{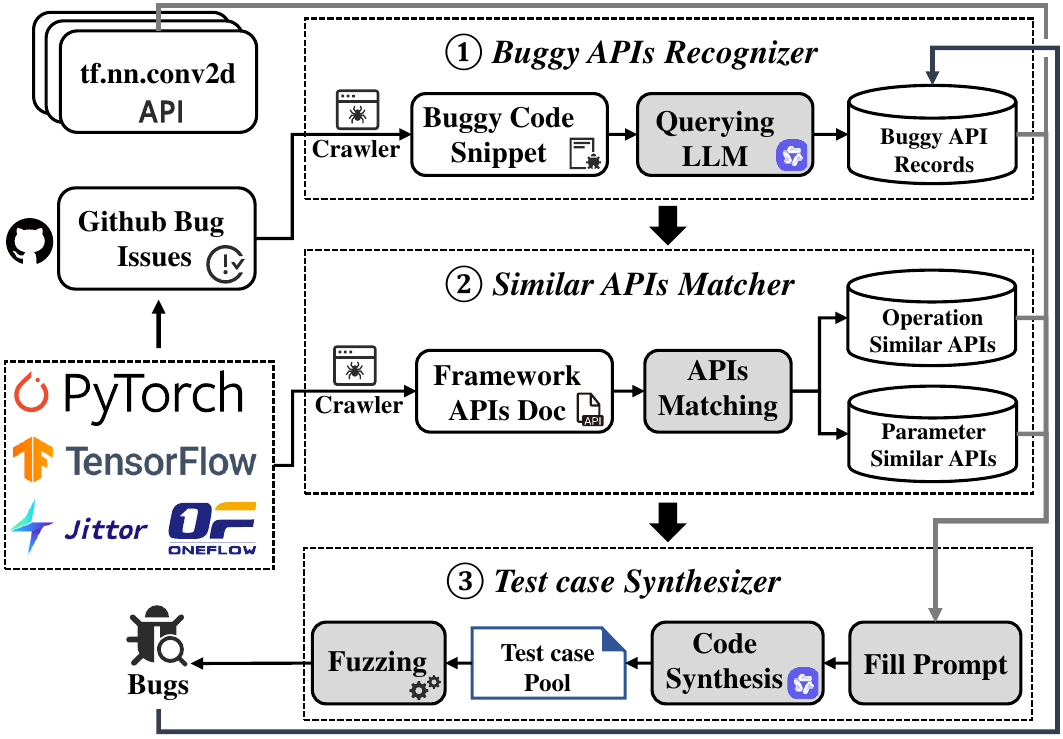}
    \caption{\sys Overview.}
    \label{fig:Overview}
\end{figure}

We design \sys to address the challenges of leveraging shared bugs for fuzzing.
Figure \ref{fig:Overview} provides an overview of our approach, which includes three major stages: recognizing buggy APIs, matching similar APIs, and synthesizing test cases for fuzzing.

In the first step, \sys uses web crawlers to collect GitHub issues related to DL frameworks. 
It filters issues based on specific bug-related keywords and employs LLM to locate buggy APIs, analyze bug causes, and identify triggering parameters.
The results are then saved in buggy API records (see Section \ref{subsection:recognizer}).
Next, \sys extracts API metadata from the documentation.
It identifies operation-similar and parameter-similar APIs within and across frameworks by measuring text and semantic similarity (see Section \ref{subsection:matcher}).
For an API under test, \sys retrieves similar APIs, queries their bug records, and includes the bug data in the LLM prompt to synthesize API invocation code (see Section \ref{subsection:synthesizer}).
Finally, \sys uses the code synthesized by the LLM as test cases for fuzz testing to enable early bug detection.
The following sections explain each step in detail.
\subsection{Buggy APIs Recognition}
\label{subsection:recognizer}
At this stage, \sys uses LLM to identify buggy APIs and triggering parameters from API-related bug reports.
It summarizes bug causes based on report context and saves the results in structured records.

We observe that CVE repositories often provide limited details about specific buggy APIs. 
However, such details are crucial for bug discovery.
For instance, PyTorch has only a few bugs documented in the CVE repository \cite{PyTorchcve}, while a significantly larger number of bug reports are actively submitted and discussed in GitHub issues \cite{PyTorchissues}.
Therefore, \sys collects API bug reports from GitHub issues.

\chadded[id=osw]{\noindent \textbf{API Bug Report Collection.} 
To be specific, we implement a web crawler based on the GitHub REST API \cite{githubrestapi} to automatically collect issues from various DL frameworks.
Since these issues include many reports unrelated to bugs (e.g., feature requests, performance improvements, help requests, and documentation issues), it is necessary to filter them.
We employ a keyword-based matching method, similar to \cite{shirihistory}, to collect bug-related issues instead of relying on the commonly used "bug" labels on GitHub. This is because such labels may not correspond to the specific types of bugs we study: some may be incorrectly assigned, labeling issues unrelated to actual bugs, while others may be missing for issues that actually cause API bugs in certain frameworks.
The keyword list is compiled using three strategies: (1) extracting common terms from prior DL framework testing studies \cite{wei2022free,xie2022docter,deng2022fuzzingDeepREL,deng2023large,shi2023acetest}, (2) adopting the custom keyword set from Orion \cite{shirihistory}, and (3) manually adding framework-specific terms, including Chinese terms found in bug reports for Jittor and OneFlow. 
The final list of selected keywords for filtering includes: ``\textit{crash}'', ``\textit{aborted (core dumped)}'', ``\textit{assertion failure}'', ``\textit{segmentation fault (core dumped)}'', ``\textit{floating point exception}'', and others.
The complete list of keywords is available in our open-source code \cite{mirrorfuzz_github} and can be extended as needed.}

\begin{table}[t]
    \centering
    \small
    \caption{Statistics of collected issues.}
    \resizebox{\columnwidth}{!}{%
    \label{tab:bug_issue_count}
    \begin{tabular}{@{}lccc@{}}
\toprule
\textbf{Framework} & \textbf{ALL Issues} & \textbf{Bug Issues} & \textbf{Code Snippets} \\ \midrule
TensorFlow & 40348 & 1835 (4.55\%) & 3693 \\
PyTorch & 46486 & 2350 (5.06\%) & 3257 \\
OneFlow & 970 & 94 (9.69\%) & 141 \\
Jittor & 356 & 57 (16.01\%) & 76 \\ \midrule
\textbf{Total} & 88160 & 4336 (4.92\%) & 7167 \\ \bottomrule
\end{tabular}}
\end{table}

As a result, we collected 88160 issues from the four frameworks on GitHub.
Using a keyword-matching approach, we identified a total of 4336 bug-related issues (refer to Table \ref{tab:bug_issue_count}).
Next, we use regular expressions to extract buggy code snippets, issue titles, and descriptions from these issues.
For example, bug code snippets are typically marked with Markdown code tags (i.e., \verb|```code```|), which allows us to extract the code based on these tags easily.
This approach may introduce false positives, as some contributors mistakenly use Markdown code blocks to represent logs, code output, etc.
Thus, these code snippets are further filtered in subsequent processing.
Moreover, we observe that many issues share reproducible bug code in cloud development environments (e.g., Google Colab \cite{google_colab}) to facilitate quick verification and debugging for developers.
To address this, we employ a web crawler to collect the relevant shared links and extract the reproducible code from them.
Eventually, we obtain structured data with the issue title, description, and code snippet.

\chadded[id=osw]{
\noindent \textbf{Buggy APIs Recognition.} 
After obtaining API-related bug issues, our next step is to identify the buggy APIs by analyzing the context of each issue.
We do not use syntax parsers for this task, because tools such as Tree-sitter \cite{tree_sitter} can only analyze syntax. They cannot interpret the semantics of bug reports containing multiple types of information (e.g., natural language descriptions, code snippets, and error logs), determine which API causes a bug, or identify the root cause.
This challenge is further compounded by the variable quality of GitHub issues, whose clarity and completeness depend on the submitters’ expertise and may lead to incomplete or unclear reports.
As a result, traditional methods such as regex matching or syntax parsers struggle to reliably extract buggy APIs and identify the root causes of bugs.}

\begin{figure}
    \centering
    \subfloat[Example prompt for buggy APIs recognition]{\includegraphics[width=0.5\textwidth]{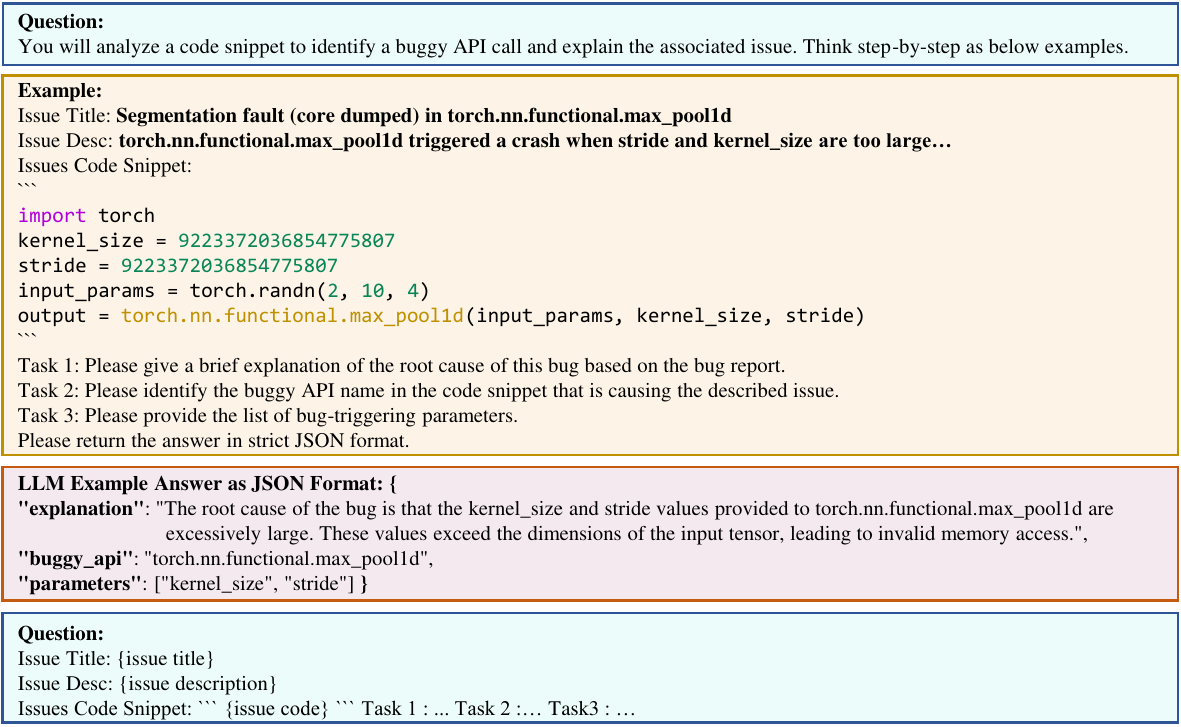}} \\
    \subfloat[Example prompt for buggy APIs verification]{\includegraphics[width=0.5\textwidth]{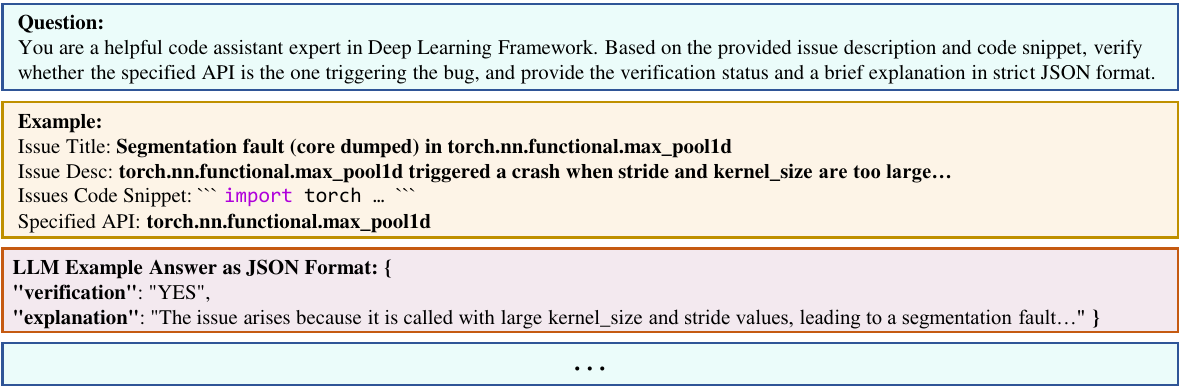}} 
    \caption{Prompt template for buggy APIs recognition and verification.}
    \label{fig:alg_locate_prompt}
\end{figure}

To overcome this, we design an automated buggy API extraction method, utilizing prompt engineering to guide LLM in recognizing buggy APIs and parameters from the issues. We observe that the code snippets in the issues often explicitly contain the buggy API. The LLM's understanding of code allows it to identify the buggy API from them.
Specifically, we first compile the code extracted from the issue to filter out non-code content that uses Markdown code tags.
Then, as shown in Figure~\ref{fig:alg_locate_prompt}, we use a few-shot Chain-of-Thought (CoT)\cite{wei2022chain} prompting technique to guide the LLM in identifying the buggy API based on the issue information and the code itself.
The LLM may generate incomplete API names based on the API calls in the code.
For example, it may identify \texttt{nn.conv2d} as the buggy API in Listing~\ref{para_sim}. 
While \texttt{nn.conv2d} is indeed a buggy API in this case, it is incomplete, as the expected name is \texttt{torch.nn.conv2d}.
To resolve this, we calculate the Levenshtein Distance \cite{Levenshtein_distance} between the complete API names and identified buggy API names to recognize the most similar API, ensuring the API name is complete. 
To further improve the accuracy of buggy API identification, we integrate the buggy APIs identified by the LLM with the issue context into a prompt, enabling the LLM to validate them.

\subsection{Similar APIs Matching}
\label{subsection:matcher}
At this stage, the matcher identifies pairs of APIs that are operation-similar and parameter-similar based on the similarity of API metadata (i.e., name, parameters, and documentation descriptions). We combine text and semantic similarity to address the challenges of matching API pairs between frameworks. Finally, to improve matching accuracy, we propose an API selection algorithm to optimize the matching results.

We first define the characteristics of these APIs and then introduce the matching process.

\chadded[id=osw]{
\noindent \textbf{Definition 1 (Operation-similar APIs)}.  
Given a pair of APIs, \(S\) and \(T\), which come from the same or different frameworks, if \(S\) and \(T\) perform the same or similar functionalities, we define them as operation-similar (OS). Formally,
\[
OS(S, T) \iff Functionality(S) \sim Functionality(T)
\]
where \({Functionality}\) denotes an API functionality and \(Functionality(S) \sim Functionality(T)\) indicates that \(S\) and \(T\) have similar functionalities.
This definition is not limited to the computational operations mentioned in the motivation, such as pooling or convolution, but also covers various categories of APIs (e.g., data processing, model building, and serialization), with a primary focus on functional similarity.
}

\noindent \textbf{Definition 2 (Parameter-similar APIs)}.  
Given a pair of APIs, \(S\) and \(T\), which come from the same or different frameworks, if they have similar parameters and are not operation-similar, we define them as parameter-similar (PS). Formally,
\begin{align*}
\text{PS}(S, T) \iff & \neg \text{OS}(S, T) \\
& \land \exists p \in \text{Params}(S) \\
& \land \exists q \in \text{Params}(T) \\
& \land p \sim q
\end{align*}
where \(\text{Params}(S)\) and \(\text{Params}(T)\) represent the sets of parameters for APIs \(S\) and \(T\), respectively, and \(p \sim q\) indicates that parameters \(p\) and \(q\) are similar.

Matching operation-similar and parameter-similar APIs within and across frameworks is challenging due to the diversity of APIs.
Fortunately, these frameworks provide documentation and API metadata, which include details such as the API name, parameters, and functionality descriptions.
For instance, with the API \texttt{torch.nn.FractionalMaxPool3d}, we can extract details from the official documentation, including its name, description (e.g., 'performs 3D fractional max pooling over an input signal consisting of multiple input planes'), and parameter information (e.g., \textit{kernel\_size}, the size of the window to take a max over).
We match API pairs by calculating the similarity of their metadata, considering both text and semantic similarities.
In particular, we define two functions, \(Sim_{text}()\) and \(Sim_{sem}()\), to measure text and semantic similarity between API \( S \) and \( T \), respectively.

\noindent \textbf{Matching Operation-similar APIs}. 
For operation-similar APIs, we first calculate text similarity based on metadata and evaluate semantic similarity and then combine them to obtain an overall similarity score.

Specifically, for a pair of APIs \( S \) and \( T \), given the complete metadata \( M^{All} \) (including API names, parameters, and descriptions), the text similarity between \( S \) and \( T \) can be calculated as follows:
\begin{equation}
\label{eq:sim_text}
\small
\begin{aligned}
   Sim_{text}(M_{S}^{All}, M_{T}^{All}) &= \max \Big( 
   Sim_{text}(M_{S}^{name},M_{T}^{name}) + \\
   &\quad Sim_{text}(M_{S}^{param},M_{T}^{param}), \\
   &\quad Sim_{text}(M_{S}^{desc},M_{T}^{desc}) + \\
   &\quad Sim_{text}(M_{S}^{param},M_{T}^{param}) \Big)
\end{aligned}
\end{equation}
Here, the \(\max\) function ensures that the highest similarity score is selected from the combined text similarity of names and parameters, or descriptions and parameters.
To calculate the text similarity \(Sim_{text}(M_{S}^{All}, M_{T}^{All})\), we use Term Frequency-Inverse Document Frequency (TF-IDF) \cite{tf_idf} and cosine similarity \cite{Cosine_similarity}. 
TF-IDF measures the importance of terms based on their frequency and the uniqueness of terms in the document collection.
In particular, TF indicates how often a term appears in a document, while IDF reduces the weight of common terms (such as ``torch'' or ``tf''), thereby emphasizing specific terms that differentiate the API. 

Formally, given a collection of documents \( D \), a document \( d \in D \), and a term \( t \), the TF-IDF calculation formula is:
\begin{equation}
\begin{aligned}
TF\textit{-}IDF(t, d, D) = TF(t, d) \times IDF(t, D)
\end{aligned}
\end{equation}
where \( TF(t, d) \) is the term frequency of \( t \) in document \( d \), and \( IDF(t, D) \) represents the inverse document frequency of term \( t \) in the document collection \( D \).
For APIs within the same framework, we construct separate documents for their name, parameters, and description metadata. 
Then, given a pair of APIs \( S \) and \( T \), specific metadata \(M\), we convert them into TF-IDF vectors \( \mathbf{v}_S^{M} \) and \( \mathbf{v}_T^{M} \), and then calculate their cosine similarity:
\begin{equation}
\begin{aligned}
Sim_{text}(M_S, M_T) = \cos(\mathbf{v}_S^{M}, \mathbf{v}_T^{M}) = \frac{\mathbf{v}_S^{M} \cdot \mathbf{v}_T^{M}}{\|\mathbf{v}_S^{M}\| \|\mathbf{v}_T^{M}\|}
\end{aligned}
\end{equation}

TF-IDF focuses solely on term occurrence frequency, overlooking semantic relationships between terms \cite{qaiser2018text}.
As a result, it considers semantically similar terms like ``GetData'' and ``FetchData'' to be completely different. 
This limitation complicates matching similar APIs across frameworks with differing naming conventions.
Thus, we introduce semantic similarity to address this limitation by capturing the semantic relationships between metadata.
Similarly, we calculate semantic similarity using the same Equation \ref{eq:sim_text} as for text similarity.
Then, we use the M3-embedding model \cite{chen2024m3} to capture semantic information from the metadata of APIs.
The M3-embedding model generates embeddings where cosine distance reflects semantic similarity.

Formally, given a text \(X\), the M3-embedding encodes it into a semantic vector \( \mathbf{e}_X \):
\[
    \mathbf{e}_X = M3\textit{-}embedding(X)
\]
For calculating the semantic similarity of metadata, given a pair of APIs \( S \) and \( T \), and specific metadata \(M\), we first use M3-embedding model to encode the API's specific metadata to obtain the corresponding embedding vectors \(\mathbf{e}_S^{M}\) and \(\mathbf{e}_T^{M}\), and then calculate the similarity between these embedding vectors using cosine similarity:
\begin{equation}
\begin{aligned}
Sim_{sem}(M_S, M_T) = \cos(\mathbf{e}_S^{M}, \mathbf{e}_T^{M}) = \frac{\mathbf{e}_S^{M} \cdot \mathbf{e}_T^{M}}{\|\mathbf{e}_S^{M}\| \|\mathbf{e}_T^{M}\|}
\end{aligned}
\end{equation}
Finally, we consider both text and semantic similarity, with the calculation for operation-similar APIs as follows:
\begin{equation}
\label{eq:sim_api_os}
{
\begin{aligned}
OS(S, T) = &\alpha \cdot Sim_{text}(M_{S}^{All}, M_{T}^{All}) \\
           &+ (1 - \alpha) \cdot Sim_{sem}(M{S}^{All}, M_{T}^{All})
\end{aligned}
}
\end{equation}
where \(\alpha\) is a weight parameter used to balance the contributions of text and semantic similarity.

\noindent \textbf{Matching Parameter-similar APIs.}
We also calculate text and semantic similarity to identify parameter-similar API pairs.
Note that we only match within the set of API pairs that are not operation-similar, which helps avoid redundant matches.
Similar to operation-similar APIs, we use TF-IDF to calculate text similarity and the M3-embedding model to calculate semantic similarity.
Notably, \( PS(S, T) \) relies on only one metadata, allowing direct calculation of similarity without adhering to Equation \ref{eq:sim_text}.
Therefore, given a pair of APIs \( S \) and \( T \), along with their parameter metadata \( M^{param} \), the calculation for parameter-similar APIs is as follows:
\begin{equation}
\label{eq:sim_api_ps}
{
\begin{aligned}
PS(S, T) = &\alpha \cdot Sim_{text}(M_{S}^{Param}, M_{T}^{Param}) \\
           &+ (1 - \alpha) \cdot Sim_{sem}(M_{S}^{Param}, M_{T}^{Param})
\end{aligned}
}
\end{equation}
where a weight parameter is the same as the definition of operation-similar APIs.
\begin{algorithm}[t]
\caption{API Selection Optimization}
\label{Similar_API_Selection}
\KwIn{source API \textit{S}, target API list \textit{L}, top-k \textit{K}, threshold \textit{H}}
\KwOut{list of selected similar APIs \textit{R}}

\SetKwFunction{SimAPI}{CalculateSimilarity}
\SetKwFunction{SortBySim}{SortSimDesc}

\textit{R} $\gets \{\}$\;
\textit{sim\_list} $\gets \{\}$\;

\ForEach{$T \in L$}{
    $sim\_score \gets$ \SimAPI{S, T}\;
    $sim\_list \gets sim\_list \cup \{(T, sim\_score)\}$\;
}

\textit{sim\_list} $\gets$ \SortBySim{(sim\_list)}\;
$R \gets sim\_list[0:K]$\;

\ForEach{ $(T, sim\_score) \in sim\_list[K:]$}{
    \If{sim\_score $\geq$ H}{
        $R \gets R \cup \{(T, sim\_score)\}$\;
    }
}
\KwRet{R}\;
\end{algorithm}

\noindent \textbf{API Selection Optimization.}
\chadded[id=osw]{
Each source API has a different number of similar APIs, and the distributions of their similarity scores can vary across frameworks. 
The widely used Top-$k$ method~\cite{deng2022fuzzingDeepREL,cai2019biker} has limitations: a small $k$ may miss valuable APIs, while a large $k$ increases fuzzing overhead. Similarly, a fixed threshold \textit{H} can also be problematic due to differences across frameworks.
To address these, we propose Algorithm~\ref{Similar_API_Selection}, a two-stage filtering strategy combining Top-$k$ selection with a similarity threshold \textit{H}, instead of relying solely on either Top-$k$ or threshold-based methods.
First, it calculates the text and semantic similarity between source API $S$ and all target APIs $L$ (line~4).
Then, it selects the top $k$ APIs with the highest similarity scores (line~7). 
Additionally, it applies threshold filtering to the remaining APIs. 
Those ranked $k+1$ and beyond are included if their similarity exceeds the threshold (line~9).
In summary, Top-$k$ ensures a minimum set of high-quality matches, while the threshold \textit{H} further supplements this set by including additional relevant APIs beyond $k$.
When the similarity scores of the Top-$k$ APIs are all greater than or equal to \textit{H}, the algorithm effectively reduces to a purely threshold-based strategy.
We set the default Top-$k$ value to 6, which may occasionally include a few irrelevant APIs. However, the powerful reasoning capabilities of modern LLMs can handle these cases, and the additional overhead is considered acceptable in the overall fuzzing process.
The threshold \textit{H} is determined through empirical observation and is primarily influenced by the scope of API matching. Specifically, we assign different thresholds for within-framework and cross-framework matching, since even functionally similar APIs across frameworks can differ significantly in naming conventions and documentation. Notably, the threshold \textit{H} and the value of $k$ are not fixed and can be adjusted by users or researchers to suit different requirements and practical scenarios.
}

\subsection{Test Case Synthesis}
\label{subsection:synthesizer}
In this stage, MirrorFuzz generates test cases for the API under test by leveraging an LLM to analyze the root causes of bugs in similar APIs. 
These test cases are carefully crafted to expose analogous bugs in the API under test.
Then, they are stored in the initial test case pool for subsequent fuzz testing.

\chadded[id=osw]{
The process of code synthesis is critical, as it must account for the complex input constraints and parameter dependencies of various framework APIs. Although bug patterns may be shared, the test code itself cannot be directly reused. This is because even functionally similar APIs may differ in naming conventions, parameter specifications, or framework-specific constructs. Bugs may also arise from diverse root causes, including invalid API call sequences, tensor shape mismatches, or data type incompatibilities. As a result, simply migrating the inputs that caused a bug in one API may produce invalid test cases and fail to trigger bugs in the target API.}

\chadded[id=osw]{
Recent studies \cite{deng2023large, deng2024largefuzzgpt} have demonstrated that LLMs, trained on extensive corpora containing code snippets for DL framework APIs, can effectively understand the syntax and semantics of these APIs. 
This capability enables LLMs to generate valid API invocation codes suited to specific DL frameworks.
Therefore, MirrorFuzz leverages the reasoning and code generation capabilities of LLMs to synthesize code, which is then used as test cases for fuzz testing.
To achieve this, MirrorFuzz retrieves buggy API records from APIs that are operation-similar and parameter-similar to the API under test.
These records include details such as buggy APIs, problematic parameters, relevant code snippets, and the root causes of the bugs.
Using this information, MirrorFuzz guides the LLM to create test cases designed to trigger similar bugs in the API under test.}

\begin{figure}
    \centering
    \includegraphics[width=1\linewidth]{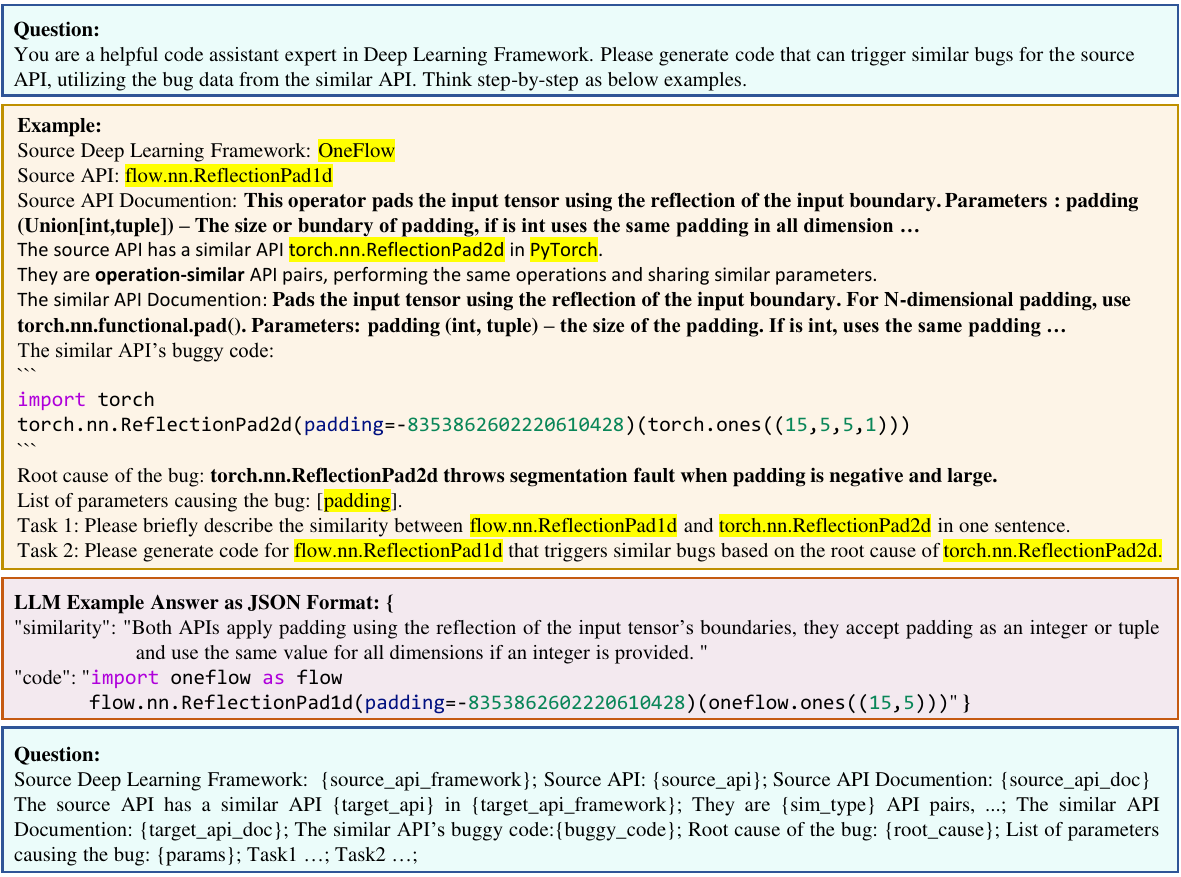}
    \caption{Prompt template for test case synthesis.}
    \label{fig:synthesis_prompt}
\end{figure}

\chadded[id=osw]{\noindent \textbf{Code Synthesis.}
Specifically, the code synthesis process is as described in Algorithm \ref{alg:shared_bug_synthesis}.
We first query APIs that are similar in operations and parameters to the API \(S\), as well as the documentation information for API \(S\), and initialize \(T\) to store the test cases (line~1-4).
Next, we iterate through the list of APIs with operations and parameters similar to API \(S\) and query their bug data from the bug records (line~5-7). 
After querying the bug data, we design a prompt (i.e., \(syn\_prompt\)) for code synthesis, allowing the LLM to generate test cases for API \(S\) based on the queried bug data (line~9-10).
We execute the synthesized code to verify that the test cases are executable (line~11).
If syntax or runtime errors occur, we use the LLM to attempt fixes based on the error messages (line~12-15).
This process deals with exceptions in LLM-synthesized code.
When a synthesized test case fails due to syntax or runtime errors (e.g., violations of API constraints), we provide the LLM with the exception information to repair it, giving the test case additional chances to meet syntactic and semantic requirements while preserving its potential to expose shared bugs.
Finally, we evaluate the quality of the test cases and save them into the test case pool \(T\) (line~16-17).
For \(syn\_prompt\), we employ the few-shot CoT prompting technique, as illustrated in Figure~\ref{fig:synthesis_prompt}.}

\begin{algorithm}[t]
\newcommand{\commentcode}[1]{\textcolor{blue}{\scriptsize #1}}
\caption{Test case Synthesis}
\small
\label{alg:shared_bug_synthesis}
    \KwIn{API under test \textit{S}}
    \KwOut{test case pool for the API under test \textit{T} }
    
    \SetKwFunction{FPromptConstruction}{PromptConstruction}
    \SetKwFunction{FMain}{CodeSynthesis}
    \SetKwFunction{QueryOperationSim}{QueryOperationSimAPIs}
    \SetKwFunction{QueryParameterSim}{QueryParameterSimAPIs}
    \SetKwFunction{FillPrompt}{FillPrompt}
    \SetKwFunction{LoadBuggyAPIDB}{LoadBuggyAPIDB}
    \SetKwFunction{QueryAPIDoc}{QueryAPIDoc}
    \SetKwFunction{QueryBugAPIDB}{QueryBugRecords}
    \SetKwFunction{UpdateBugAPIDB}{UpdateBugRecords}
    \SetKwFunction{FQueryLLM}{LLM}
    \SetKwFunction{Fuzzer}{Fuzzer}
    \SetKwFunction{CheckStatus}{CheckStatus}
    \SetKwFunction{RecordCrash}{RecordCrash}
    \SetKwFunction{exec}{exec}
    \SetKwFunction{FixbyLLM}{FixbyLLM}
    \SetKwFunction{GenbyLLM}{GenbyLLM}
    \SetKwFunction{Priority}{Priority}
    
    \SetKwProg{Fn}{Function}{:}{}
    $oper\_list \gets \QueryOperationSim{S}$\;
    $para\_list \gets \QueryParameterSim{S}$\;
    $S\_doc \gets \QueryAPIDoc{S}$\;  
    $T \gets \{\}$\;
    \ForEach{$sim\_api, sim\_type \in (oper\_list, para\_list)$}{
    $bug\_list \gets \QueryBugAPIDB{S, sim\_api}$\;

    \ForEach{$bug\_data \in bug\_list$}{
  
        \commentcode{// bug\_data includes buggy APIs, parameters, code, and root cause.}\\
        $syn\_prompt \gets \FillPrompt{S, S\_doc, bug\_data, sim\_type}$\;
        
        $test\_case \gets \FQueryLLM{syn\_prompt}$\;
        $status, out \gets \exec{test\_case}$\;
        
        \If{$status \in \textup{\textsf{Error}}$}{
    
            $f\_status \gets \FixbyLLM{test\_case, status, out}$\;
            \If{$f\_status \ == \textup{\textsf{False}}$}{ 
                \textbf{continue}\;
            }
        }
      
        $score \gets \Priority{test\_case}$\;
        $T \gets T \cup \{(test\_case, score)\}$\;
           
        }
    }
    \KwRet{T};\
\end{algorithm}

\noindent \textbf{Fuzzing Executor.}
Then, \sys selects test cases from the test case pool, applies mutation operations, and executes them to detect potential bugs in DL frameworks. 
It identifies bugs by monitoring crashes within the framework.
Furthermore, for newly discovered bugs during fuzz testing, \sys updates its bug records to identify more potential shared bugs across APIs.

For test case selection, we rank each synthesized test case \( T \) for API \( S \) using the following formula:
\begin{equation}
\begin{aligned}
\text{Priority}(T) = G \cdot(U - C)
\end{aligned}
\end{equation}
Here, \( T \) represents a synthesized test case, and \(\text{Priority}(T)\) quantifies its ranking score. 
The binary flag \( G \) ensures that the generated code includes the API \( S \), preventing the hallucination issue in LLMs that may generate invalid test cases \cite{huang2023survey}.
\( U \) represents the number of unique APIs covered by the test case, aiming to explore diverse API combinations.
\( C \) represents execution time, which helps prevent long-running test cases from consuming excessive resources.
We select the top 10 test cases with the highest priority scores for subsequent mutation operations.

For the mutation strategy, we randomly select parameters of the API under test and apply three types of mutation operators until the desired number of test cases is generated.
These operators include:
 1) boundary value mutation, which randomly inserts boundary values (e.g,. NaN, negative values, extreme numerical values) into the input data;
2) type mutation, which randomly changes the data types of parameters, such as converting \textit{tf.int32} to \textit{tf.float16}; and 3) shape and dimension mutation, which modifies input shapes and sizes, such as reshaping tensors.

For test oracle, we detect bugs by monitoring framework crashes, such as program aborts, segmentation faults, and floating-point exceptions.
Additionally, we observe that certain internal errors do not result in crashes but instead cause the framework to halt, often accompanied by messages such as, ``Please report this bug to the relevant platform/organization.''
Therefore, we include these cases in our monitoring process to ensure they are properly recorded and reported.
Since Jittor \cite{hu2020jittor} integrates an operator compiler and tuner, it compiles the entire framework and meta-operators just-in-time.
As a result, we also detect bugs where Jittor encounters compilation failures or internal failures.

\section{EVALUATION}
\label{section:Implementation}

\subsection{Implementation}
We implement \sys in Python and select commonly used DL Frameworks, including PyTorch (v2.4.1) \cite{paszke2019PyTorch}, TensorFlow (2.17.0) \cite{abadi2016tensorflow}, OneFlow (v0.9) \cite{OneFlowgithub}, and Jittor (v1.3.9.5) \cite{hu2020jittor}, all of which have been previously studied in DL framework testing research \cite{deng2024largefuzzgpt, yang2023fuzzing, liu2023generation}.
It utilizes the bs4 library \cite{pypibs4} to parse the API documentation of DL frameworks, supporting the API information needs of its three modules.
The buggy API recognition module is developed using the requests library \cite{github_requests} to scrape GitHub issues.
The similar API matching module is implemented based on M3-embedding \cite{chen2024m3}, which is employed to obtain semantic embeddings for API pairs.
We set \(\alpha\) to 0.35, prioritizing semantic similarity to address the challenges of cross-framework matching.
We then retain the Top-6 similar APIs.
For the LLM, we utilize Qwen2.5-Coder-7B-Instruct-AWQ \cite{hui2024qwen2} for test case synthesis.
Qwen2.5-Coder represents the latest series of code-specific Qwen LLMs, offering significant improvements in code generation, code reasoning, and code fixing.
We obtain the model files from Hugging Face \cite{HuggingFace} and deploy them locally.

\subsection{Experimental Setup}
\noindent \textbf{Baseline Selection.} 
Our approach is an API-level fuzzer, so we focus on state-of-the-art API-level fuzzers, including TENSORSCOPE\cite{deng2023differential}, TitanFuzz\cite{deng2023large}, Orion\cite{shirihistory}, FuzzGPT\cite{deng2024largefuzzgpt}, and Future\cite{li2024seeds}.
To gain a comprehensive understanding, we also consider model-level fuzzers such as MoCo\cite{ji2024moco}.
Among these, FuzzGPT and Future are promised to be open-source. However, as of the completion of our work, their code is still not publicly available. 
Therefore, we choose TENSORSCOPE, TitanFuzz, Orion, and MoCo as our baselines.

\noindent \textbf{DL Framework.}
Since some baselines do not support OneFlow and Jittor, we mainly conduct evaluations on TensorFlow and PyTorch.
Except for the DL Framework versions implemented in MirrorFuzz, we also conduct bug detection for each baseline on PyTorch (v1.31.1) and TensorFlow (2.10.0) to ensure a fair comparison, as these versions are the same as or closest to those used by most baselines, minimizing version differences.

\noindent \textbf{Large Language Models.} 
To comprehensively evaluate the performance of MirrorFuzz in synthesizing shared bugs using LLMs, we selected Qwen2.5-coder\cite{hui2024qwen2}, CodeGemma\cite{team2024codegemma}, and CodeLlama\cite{roziere2023codellama}, deploying them in a local environment.
The specifications of the LLMs used in our evaluation are listed in Table \ref{LLMs_evaluate}.

\begin{table}[t]
\caption{LLMs evaluated in \sys.}
\setlength{\tabcolsep}{0.4mm}
\label{LLMs_evaluate}
\resizebox{\columnwidth}{!}{
\begin{tabular}{@{}lcccc@{}}
\toprule
\textbf{LLM} & \multicolumn{1}{c}{\textbf{Model}} & \multicolumn{1}{c}{\textbf{Quantization}} & \multicolumn{1}{c}{\textbf{Generation Speed}} & \multicolumn{1}{c}{\textbf{Size}} \\ \midrule
CodeLLama & 7B-Instruct & 4-bit & 15.03 tokens/s & 3.89GB \\ 
CodeGemma & 7B-IT(Instruct Tuned) & 4-bit & 12.23 tokens/s & 6.68GB \\
Qwen2.5-Coder & 7B-Instrcut & 4-bit & 13.05 tokens/s & 5.57GB \\ \bottomrule
\end{tabular}}
\end{table}

\begin{table*}[t]
\centering
\caption{Experimental results of buggy APIs recognition in different prompt strategies.}
\label{Buggy_APIs_Identification}
\resizebox{0.95\textwidth}{!}{
\begin{tabular}{@{}l|r|c|cc|cc|cc|cc@{}}
\toprule
\multirow{2}{*}{\textbf{Framework}} & \multicolumn{1}{l|}{\multirow{2}{*}{\textbf{Codes}}} & \multirow{2}{*}{\textbf{Approach}} & \multicolumn{2}{c|}{\textbf{\#Issue w/o T}} & \multicolumn{2}{c|}{\textbf{\#Issue w/o D}} & \multicolumn{2}{c|}{\textbf{\#Issue w/o TD}} & \multicolumn{2}{c}{\textbf{\#Issue w/ ALL}} \\ \cmidrule(l){4-11} 
 & \multicolumn{1}{l|}{} &  & \textbf{APIs Acc} & \textbf{Param Acc} & \multicolumn{1}{l}{\textbf{APIs Acc}} & \multicolumn{1}{l|}{\textbf{Param Acc}} & \textbf{APIs Acc} & \textbf{Param Acc} & \textbf{APIs Acc} & \textbf{Param Acc} \\ \midrule
\multirow{2}{*}{\textbf{TensorFlow}} & \multirow{2}{*}{100} & Zero-shot & 91.20\% & 59.30\% & 92.80\% & 58.80\% & 90.80\% & 49.80\% & 93.80\% & 64.60\% \\
 &  & Few-shot CoT & 96.50\% & 63.00\% & 95.30\% & 60.70\% & 93.80\% & 55.60\% & \textbf{96.70\%} & \textbf{66.80\%} \\ \midrule
\multirow{2}{*}{\textbf{PyTorch}} & \multirow{2}{*}{100} & Zero-shot & 75.00\% & 65.60\% & 79.90\% & 66.90\% & 69.90\% & 54.70\% & 80.70\% & 68.30\% \\
 &  & Few-shot CoT & 87.40\% & 68.30\% & 91.40\% & 69.20\% & 83.40\% & 63.20\% & \textbf{92.10\%} & \textbf{71.00\%} \\ \midrule
\multirow{2}{*}{\textbf{OneFlow}} & \multirow{2}{*}{50} & Zero-shot & 90.80\% & 76.70\% & 91.00\% & 72.70\% & 88.40\% & 66.90\% & 91.80\% & 77.80\% \\
 &  & Few-shot CoT & 92.20\% & \textbf{78.60\%} & \textbf{94.20\%} & 72.90\% & 92.20\% & 68.60\% & 92.40\% & 75.50\% \\ \midrule
\multirow{2}{*}{\textbf{Jittor}} & \multirow{2}{*}{30} & Zero-shot & 94.00\% & 88.00\% & 99.00\% & 90.00\% & 99.30\% & 93.30\% & 90.00\% & 87.30\% \\
 &  & Few-shot CoT & 99.00\% & 89.30\% & 99.70\% & \textbf{94.70\%} & 99.70\% & 91.70\% & \textbf{99.70\%} & 90.00\% \\ \midrule
\multirow{2}{*}{\textbf{Total}} & \multirow{2}{*}{280} & Zero-shot & 87.75\% & 72.40\% & 90.68\% & 72.10\% & 87.10\% & 66.18\% & 89.08\% & 74.50\% \\
 &  & Few-shot CoT & 93.78\% & 74.80\% & 95.15\% & 74.38\% & 92.28\% & 69.78\% & \textbf{95.23\%} & \textbf{75.83\%} \\ \bottomrule
\end{tabular}}
\end{table*}

\noindent \textbf{Environment.} We use a 64-core workstation with 256 GB RAM, running Ubuntu 22.04 LTS, and equipped with 2 NVIDIA RTX 4090 GPUs.

\subsection{Evaluation Metrics.} 
\noindent \textbf{Number of covered APIs.} We report the number of API coverages, as there are a large number of APIs in the DL framework.

\noindent \textbf{Code coverage.}
Code coverage is a widely used test adequacy metric in testing DL frameworks.
We use the coverage.py \cite{pythoncoverage} library to report the number of covered Python code lines in the DL framework.

\noindent \textbf{Number of detected bugs.}
Following prior work on fuzzing DL Frameworks \cite{wei2022free,deng2022fuzzingDeepREL,deng2023large,deng2024largefuzzgpt}, we report the number of detected bugs.

\noindent \textbf{Validity and success rate of synthesis.}
MirrorFuzz leverages LLMs to synthesize initial test cases.
We evaluate the synthesis validity rate via JSON parsing and Python syntax detection, as well as the success rate of LLMs in reproducing known bugs in similar APIs.


\section{RESULTS ANALYSIS}
\label{section:Evaluation}
We evaluate MirrorFuzz by answering the following four research questions:
\begin{itemize}[left=0pt]
    \item  \textbf{RQ1:} How do different prompt strategies affect the accuracy of recognizing buggy APIs from bug reports?

    \item  \textbf{RQ2:} How prevalent are similar APIs across DL frameworks, and how effectively can \sys identify them?
    
    \item \textbf{RQ3:} How effective is MirrorFuzz in synthesizing shared bugs and detecting real-world bugs?
    
    \item \textbf{RQ4:} How does \sys perform relative to state-of-the-art approaches?

\end{itemize}

\subsection{Accuracy of Buggy APIs Recognition}
In this section, we conduct an ablation study to evaluate the effectiveness of the prompting strategy for buggy API recognition.
We randomly sample 280 issues from the crawled GitHub issues for four frameworks, with the three authors manually identifying buggy APIs and bug-causing parameters, followed by cross-validation to ensure accuracy.
We design four prompt variants: ``\#Issue w/o T'' (without issue title), ``\#Issue w/o D'' (without description), ``\#Issue w/o TD'' (without both), and ``\#Issue w/ ALL'' (with code, title, and description).
These variants are evaluated under two conditions: few-shot CoT and zero-shot, using the latter as the baseline.
To mitigate LLM randomness, we repeat each experiment independently 10 times per framework and average the results.
As shown in Table \ref{Buggy_APIs_Identification}, providing buggy API recognition examples, complete issue context, and refining the task through few-shot CoT techniques improves the LLM’s performance, achieving 95.23\% accuracy in buggy API recognition and 75.83\% in bug-causing parameter identification. 
This demonstrates that our method effectively identifies buggy APIs and bug-causing parameters from GitHub issues and is effective across multiple frameworks.

Furthermore, we observe that in OneFlow, ``\#Issue w/o T'' achieves the highest accuracy in bug-causing parameter recognition, while ``\#Issue w/o D'' excels in buggy API recognition.
In Jittor, ``\#Issue w/o D'' achieves the highest accuracy in bug-causing parameter recognition.
This may stem from the fact that issue reports submitted in OneFlow and Jittor often fail to meet community standards, resulting in generally lower-quality reports compared to those of TensorFlow and PyTorch.
For instance, many descriptions are vague and lack clarity, failing to provide the root causes of the reported issues. 
Such issues can hinder LLMs' ability to analyze and handle them effectively.

\begin{table}[t]
\centering
\caption{Matched operation-similar (\#OS) and parameter-similar (\#OP) API pairs within and across DL Frameworks.}
\label{matched_api_pair_os_op}
\large
\resizebox{0.95\columnwidth}{!}{
\begin{tabular}{@{}l|rr|rr|rr|rr@{}}
\toprule
 & \multicolumn{2}{c|}{\textbf{TensorFlow}} & \multicolumn{2}{c|}{\textbf{PyTorch}} & \multicolumn{2}{c|}{\textbf{OneFlow}} & \multicolumn{2}{c}{\textbf{Jittor}} \\ \cmidrule(l){2-9} 
 & \multicolumn{1}{c}{\textbf{\#OS}} & \multicolumn{1}{c|}{\textbf{\#PS}} & \multicolumn{1}{c}{\textbf{\#OS}} & \multicolumn{1}{c|}{\textbf{\#PS}} & \multicolumn{1}{c}{\textbf{\#OS}} & \multicolumn{1}{c|}{\textbf{\#PS}} & \multicolumn{1}{c}{\textbf{\#OS}} & \multicolumn{1}{c}{\textbf{\#PS}} \\ \midrule
\textbf{TensorFlow} & 19980 & 7467 & 2408 & 737 & 1277 & 227 & 800 & 420 \\
\textbf{PyTorch} & 2509 & 609 & 4502 & 2355 & 6310 & 937 & 2252 & 807 \\
\textbf{OneFlow} & 1269 & 73 & 6063 & 571 & 714 & 144 & 623 & 153 \\
\textbf{Jittor} & 808 & 539 & 2275 & 841 & 628 & 234 & 1357 & 178 \\ \midrule
\textbf{Total} & 24566 & 8688 & 15248 & 4504 & 8929 & 1542 & 5032 & 1558 \\ \bottomrule
\end{tabular}}
\end{table}

\begin{table}[t]
\caption{Evaluation of API matching methods under different settings.}
\label{matcher_cov_rate}
\centering
\small
\resizebox{0.95\columnwidth}{!}{
\begin{tabular}{@{}l|rr|rr@{}}
\toprule
\textbf{} & \multicolumn{2}{c|}{\textbf{Matcher w/o optimization }} & \multicolumn{2}{c}{\textbf{Matcher w/ optimization }} \\ \cmidrule(l){2-5} 
 & \multicolumn{1}{c}{\textbf{\#Within}} & \textbf{\#Cross} & \multicolumn{1}{c}{\textbf{\#Within}} & \textbf{\#Cross} \\ \midrule
\textbf{TensorFlow} & 63.52\% & 89.79\% & 80.85\% & 91.27\% \\
\textbf{PyTorch} & 66.33\% & 78.39\% & 82.99\% & 89.90\% \\
\textbf{OneFlow} & 85.94\% & 70.60\% & 89.06\% & 76.54\% \\
\textbf{Jittor} & 88.69\% & 81.48\% & 92.26\% & 86.48\% \\ \midrule
\textbf{Total} & 76.12\% & 80.07\% & \textbf{86.29\%} & \textbf{86.05\%} \\ \bottomrule
\end{tabular}}
\end{table}

\subsection{Extent of API Similarity in DL Frameworks}
To uncover API commonalities across frameworks, we use our method to quantify operation-similar (\#OS) and parameter-similar (\#PS) APIs within and across frameworks. As shown in Table \ref{matched_api_pair_os_op}, each row and column represents a framework, with rows as the source and columns as the target.
For example, TensorFlow has 2408 \#OS and 737 \#PS API pairs in PyTorch, while PyTorch has 2509 and 609 in TensorFlow.
The table reveals numerous similar API pairs across frameworks, indicating that frameworks are widely vulnerable to shared bugs.
Notably, TensorFlow and PyTorch have the most matching API pairs, primarily because they offer more APIs than Jittor and OneFlow.
Overall, \#PS API pairs are fewer than \#OS ones.
However, APIs with different names and functionalities may share parameter designs, potentially causing shared bugs.
Thus, \#PS pairs are a crucial complement to \#OS ones.

\chadded[id=osw]{
Additionally, we randomly select 40 APIs (10 from each of four frameworks) and ask three authors familiar with DL frameworks to manually label similar API pairs within and across frameworks.
To evaluate the effectiveness of our method, we calculate the coverage rate of the labeled API pairs identified by our method.
In total, we manually label 596 \#OS API pairs for 40 APIs, including 156 within-framework (``\#Within'') pairs and 440 cross-framework (``\#Cross'') pairs. 
We conduct experiments both with and without API selection optimization (Algorithm \ref{Similar_API_Selection}). ``Matcher w/ optimization'' uses Top-$k$ selection combined with a similarity threshold, while ``Matcher w/o optimization'' uses only Top-$k$ selection.
As shown in Table \ref{matcher_cov_rate}, the default setting ``Matcher w/ optimization'' achieves average matching coverage rates of 86.29\% within frameworks and 86.05\% across frameworks, outperforming the ``Matcher w/o optimization'' setting. 
This demonstrates that our method can effectively match similar APIs both within and across frameworks, and that the API selection optimization further enhances the results.}

\begin{table*}[t]
\caption{Synthesis validity and success rates of shared bugs. 
\#OS denotes operation-similar, and \#PS denotes parameter-similar.}
\label{table:Synthesis_rate}
\centering
\resizebox{0.95\textwidth}{!}{
\begin{tabular}{@{}l|c|ccc|ccc@{}}
\toprule
\multirow{2}{*}{} & \multirow{2}{*}{\textbf{LLM}} & \multicolumn{3}{c|}{\textbf{Validity Rate}} & \multicolumn{3}{c}{\textbf{Success Rate}} \\ \cmidrule(l){3-8} 
 &  & \textbf{\#OS} & \textbf{\#PS} & \textbf{ALL} & \textbf{\#OS} & \textbf{\#PS} & \textbf{ALL} \\ \midrule
\multirow{3}{*}{\textbf{Zero-shot}} & CodeLLama & 66.67\%(20/30) & 33.33\%(10/30) & 50.00\%(30/60) & 40.00\%(4/10) & 30.00\%(3/10) & 35.00\%(7/20) \\
 & CodeGemma & 76.67\%(23/30) & 70.00\%(21/30) & 73.33\%(44/60) & 40.00\%(4/10) & 60.00\%(6/10) & 50.00\%(10/20) \\
 & Qwen2.5-Coder & 100.00\%(30/30) & 93.33\%(28/30) & 96.67\%(58/60) & 50.00\%(5/10) & 80.00\%(8/10) & 65.00\%(13/20) \\ \midrule
\multirow{3}{*}{\textbf{MirrorFuzz}} & CodeLLama & 73.33\%(22/30) & 56.67\%(17/30) & 65.00\%(39/60) & 60.00\%(6/10) & 40.00\%(4/10) & 50.00\%(10/20) \\
 & CodeGemma & 93.33\%(28/30) & 83.33\%(25/30) & 88.33\%(53/60) & 60.00\%(6/10) & 70.00\%(7/10) & 65.00\%(13/20) \\
 & Qwen2.5-Coder & 100.00\%(30/30) & 100.00\%(30/30) & \textbf{100.00\%(60/60)} & 100.00\%(10/10) & 70.00\%(7/10) & \textbf{85.00\%(17/20)} \\ \bottomrule
\end{tabular}}
\end{table*}

\begin{table*}[t]
\caption{Bugs found by \sys. For the bug type, ``Segv'' stands for segmentation fault, ``FPE'' is for floating point exception, and ``Abort'' means program abort. ``Other'' includes internal failures and compilation failures.}
\label{table:Bugs_found}
\centering
\resizebox{0.95\textwidth}{!}{
\begin{tabular}{@{}l|r|ccc|ccc|cccc@{}}
\toprule
\multirow{2}{*}{\textbf{}} & \multicolumn{1}{c|}{\multirow{2}{*}{\textbf{Version}}} & \multicolumn{3}{c|}{\textbf{Total Bugs}} & \multicolumn{3}{c|}{\textbf{Status}} & \multicolumn{4}{c}{\textbf{Type}} \\ \cmidrule(l){3-12} 
 & \multicolumn{1}{c|}{} & \textbf{Total} & \textbf{Unknown} & \textbf{Known} & \textbf{Confirmed} & \textbf{Rejected} & \textbf{Fixed} & \textbf{Segv} & \textbf{FPE} & \textbf{Abort} & \textbf{Other} \\ \midrule
\textbf{TensorFlow} & 2.16 \& 2.17 \& nightly & 118 & 83 & 35 & 83 & 7 & 20 & 11 & 1 & 106 & 0 \\
\textbf{PyTorch} & 2.3.0 \& nightly & 48 & 47 & 1 & 45 & 2 & 17 & 15 & 10 & 13 & 10 \\
\textbf{OneFlow} & 0.9 \& 1.0 & 86 & 77 & 9 & 28 & 0 & 28 & 12 & 2 & 72 & 0 \\
\textbf{Jittor} & 1.3.9.5 \& 1.3.9.10 & 63 & 55 & 8 & 24 & 0 & 15 & 1 & 0 & 6 & 56 \\ \midrule
\textbf{Total} & \textbf{} & \textbf{315} & \textbf{262} & \textbf{53} & \textbf{180} & \textbf{9} & \textbf{80} & \textbf{39} & \textbf{13} & \textbf{197} & \textbf{66} \\ \bottomrule
\end{tabular}}
\end{table*}

\subsection{Effectiveness of Synthesis and Detected Bugs}
\noindent \textbf{Validity and Success Rate of Synthesis.}
To evaluate the effectiveness of MirrorFuzz in synthesizing shared bugs, we constructed a benchmark consisting of 20 shared bugs: 10 for \#OP APIs and 10 for \#PS APIs, covering 4 frameworks.
We use a zero-shot approach as the baseline, where LLMs directly synthesize shared bugs based on the provided API pairs and bug code. Each model has three opportunities to synthesize.
Table \ref{table:Synthesis_rate} shows the validity rate and success rate of synthesizing shared bugs for MirrorFuzz and the baseline.
The few-shot CoT prompt, constructed by MirrorFuzz, achieves a 100\% validity rate for the synthesized shared bugs under Qwen2.5-Coder.
This means that the generated code passes both JSON parsing and Python syntax checks.
Moreover, it successfully reproduces 17 shared bugs, achieving a 100\% success rate for \#OS APIs, outperforming both the baseline and other models.
This demonstrates that our method enables LLMs to effectively synthesize shared bugs.
Furthermore, Qwen2.5-Coder exhibits superior code understanding and generation capabilities, further enhancing its performance. Therefore, \sys uses Qwen2.5-Coder by default.

\chadded[id=osw]{\noindent \textbf{Real-world Bugs Detected.}
Table \ref{table:Bugs_found} summarizes real-world bugs detected by \sys across four popular DL frameworks.
The ``Total'' column indicates the total number of bugs detected by \sys, while the ``Unknown'' and ``Known'' columns show the number of previously unknown and previously known bugs, respectively. 
Following \cite{deng2022fuzzingDeepREL, yang2023fuzzing}, we report the number of bugs confirmed by developers as previously unknown (``Confirmed'' column); the number of bugs rejected by developers (``Rejected'' column); and the number of previously unknown bugs that have been fixed (``Fixed'' column).
Additionally, we categorize bugs by type, including segmentation faults (``Segv''), floating-point exceptions (``FPE''), program aborts (``Abort''), and framework internal failures (``Other'').}

\chadded[id=osw]{Overall, \sys detects 315 bugs, of which 262 are previously unknown. Among these, 180 are confirmed by developers, 80 are fixed, and 9 are rejected. 
Among the rejected bugs, 4 are due to developers being unable to reproduce our proof-of-concept, 3 are due to incorrect or unsupported API usage (classified by developers as user errors rather than framework bugs), and 2, though reported for the first time, cannot be reproduced in the latest nightly builds.
Across all bug types, we observe 39 segmentation faults, 13 floating-point exceptions, 197 program aborts, and 66 framework internal errors.
Most bugs in Jittor fall under the ``Other'' category, such as compilation and internal failures, due to its just-in-time compilation.}

\chadded[id=osw]{
We further evaluate bug detection capability by enabling AddressSanitizer (ASAN)~\cite{serebryany2012addresssanitizer} on TensorFlow and PyTorch, while builds for OneFlow and Jittor fail due to compilation issues. Using ASAN, we identify 9 \texttt{NULL}-pointer dereferences (1 memory write, 8 memory reads) and 1 heap buffer overflow in TensorFlow, whereas in PyTorch we find 8 \texttt{NULL}-pointer dereferences (2 memory writes, 6 memory reads), 13 heap buffer overflows, 1 container overflow, and 1 stack overflow. These findings demonstrate that MirrorFuzz is capable of detecting multiple types of bugs, including out-of-bounds reads, which may lead to severe consequences such as memory corruption or denial of service.
Furthermore, 52 bugs are assigned CNVD IDs, and 5 bugs are marked as high-priority by PyTorch developers for urgent resolution.}

\noindent \textbf{Source Distribution of Bugs.}
As shown in Table \ref{table:bugs_distribution}, we manually analyze the bugs detected by \sys to identify their origins, specifically which frameworks' operation-similar (\#OS) and parameter-similar (\#PS) APIs are responsible.
We observe that bugs in these frameworks affect each other, and most of these bugs are influenced by flaws in TensorFlow and PyTorch.
This is mainly due to TensorFlow and PyTorch being more comprehensive frameworks with more APIs and historical issues.
Interestingly, bugs in Jittor and OneFlow can also impact TensorFlow and PyTorch, despite having fewer APIs and known issues.
This highlights that even smaller or newer frameworks can impact mainstream ones, emphasizing the prevalence of shared bugs.
Thus, we should focus not only on bugs within individual frameworks but also on threats arising between framework similarities.
Additionally, \#OS APIs are the primary source of bugs, while \#PS APIs supplement gaps that \#OS APIs cannot cover. 
Moreover, we find that some bugs are repeatedly triggered by shared bugs across different frameworks, possibly because developers make the same mistakes when implementing these features.
\begin{table}[t]
\caption{Source distribution of bugs. TF stands for TensorFlow, PT for PyTorch, OF for OneFlow, and JT for Jittor.}
\label{table:bugs_distribution}
\centering
\setlength{\tabcolsep}{1.2mm}
\resizebox{1\columnwidth}{!}{
\begin{tabular}{@{}l|c|cc|cc|cc|cc|ccc@{}}
\toprule
 & \multirow{2}{*}{\textbf{Bugs}} & \multicolumn{2}{c|}{\textbf{TF}} & \multicolumn{2}{c|}{\textbf{PT}} & \multicolumn{2}{c|}{\textbf{OF}} & \multicolumn{2}{c|}{\textbf{JT}} & \multicolumn{3}{c}{\textbf{Total}} \\ \cmidrule(l){3-13} 
 &  & \textbf{\#OS} & \textbf{\#PS} & \textbf{\#OS} & \textbf{\#PS} & \textbf{\#OS} & \textbf{\#PS} & \textbf{\#OS} & \textbf{\#PS} & \textbf{\#OS} & \textbf{\#PS} & \multicolumn{1}{l}{\textbf{Total}} \\ \midrule
\textbf{TF} & 118 & 76 & 32 & 9 & 4 & 7 & 3 & 8 & 1 & 100 & 40 & 140 \\
\textbf{PT} & 48 & 15 & 5 & 25 & 6 & 4 & 1 & 6 & 3 & 50 & 15 & 65 \\
\textbf{OF} & 86 & 34 & 8 & 25 & 9 & 22 & 9 & 27 & 5 & 108 & 31 & 139 \\
\textbf{JT} & 63 & 43 & 10 & 38 & 9 & 28 & 5 & 26 & 9 & 135 & 33 & 168 \\ \midrule
\textbf{Total} & 315 & 168 & 55 & 97 & 28 & 61 & 18 & 67 & 18 & 393 & 119 & 512 \\ \bottomrule
\end{tabular}}
\end{table}
\noindent \textbf{Case Study}. 
In the following paragraphs, we discuss bugs from \#OS and \#PS APIs, missed by previous fuzzers but found by \sys.

\noindent \textbf{Shared Bug in Operation Similarity.}
\texttt{torch.nn.Conv2d} is a fundamental component for constructing deep neural networks. Before PyTorch version 1.2, an issue existed where an improper configuration of the \texttt{stride} parameter could cause the framework to crash. This issue was later resolved. However, by leveraging the \#OS relationships between these APIs, MirrorFuzz successfully synthesized code for the \texttt{torch.ao.nn.international.quantized.Conv-\\ReLU} and \texttt{torch.ao.nn.quantized.Conv} series of APIs, revealing similar crash-inducing bugs. Developers responded promptly and addressed the reported issue.

\noindent \textbf{Shared Bug in Parameter Similarity.}
In TensorFlow, various optimizers are provided for users to choose from, helping to adjust and update model parameters, such as \texttt{tf.raw\_ops.ResourceApplyAdam} and \texttt{tf.raw\_ops.ResourceSparseApplyFtrlV2}.
These APIs lack data type checks for the \textit{accumulator} parameters, which can lead to runtime crashes.
\sys identifies one such crash and updates it in the bug records.
Subsequently, by leveraging the relationship of \#OS between their APIs, after multiple fuzzing iterations, we discover 29 similar optimizer bugs in TensorFlow.

\subsection{Comparison with Prior Work}
\begin{table}[t]
\caption{Comparison of API and code coverage between TENSORSCOPE, Orion, TitanFuzz, and \sys.}
\label{api_and_code_cov}
\centering
\resizebox{0.95\columnwidth}{!}{
\begin{tabular}{@{}l|cc|cc@{}}
\toprule
\multirow{2}{*}{} & \multicolumn{2}{c|}{\textbf{TensorFlow}} & \multicolumn{2}{c}{\textbf{PyTorch}} \\ \cmidrule(l){2-5} 
 & \textbf{API Cov} & \textbf{Code Cov} & \textbf{API Cov} & \textbf{Code Cov} \\ \midrule
\textbf{Total} & 4403 & 284510(100\%) & 1992 & 275822(100\%) \\ \midrule
\textbf{TENSORSCOPE} & 1469 & 81202(28.54\%) & - & - \\
\textbf{Orion} & \textbf{4037} & 84047(29.54\%) & 1751 & 27130(9.84\%) \\
\textbf{TitanFuzz} & 2215 & 92670(32.57\%) & 1329 & 39207(14.21\%) \\
\textbf{MirrorFuzz} & 3995 & \textbf{129663(45.57\%)} & \textbf{1868} & \textbf{77707(28.17\%)} \\ \bottomrule
\end{tabular}}
\end{table}

\begin{table}[t]
\setlength{\tabcolsep}{0.4mm}
\caption{Comparison of API and code coverage between MoCo and \sys based on the APIs supported by MoCo.}
\label{api_and_code_cov_moco}
\resizebox{\columnwidth}{!}{
\begin{tabular}{@{}l|cc|cr|cc@{}}
\toprule
\multirow{2}{*}{} & \multicolumn{2}{c|}{\textbf{TensorFlow}} & \multicolumn{2}{c|}{\textbf{PyTorch}} & \multicolumn{2}{c}{\textbf{Jittor}} \\ \cmidrule(l){2-7} 
 & \textbf{API Cov} & \textbf{Code Cov} & \textbf{API Cov} & \multicolumn{1}{c|}{\textbf{Code Cov}} & \textbf{API Cov} & \textbf{Code Cov} \\ \midrule
\textbf{Total} & 169 & 284510(100\%) & 144 & 275822(100\%) & 72 & 13166(100\%) \\ \midrule
\textbf{MoCo} & 95 & 74756(26.27\%) & 72 & 26271(9.52\%) & 66 & 3955(30.04\%) \\
\textbf{MirrorFuzz} & \textbf{145} & \textbf{83875(29.48\%)} & \textbf{142} & \textbf{67322(24.40\%)} & \textbf{68} & \textbf{4841(36.77\%)} \\ \bottomrule
\end{tabular}}
\end{table}

We compare \sys with state-of-the-art API-level and model-level fuzzers.
Despite our best efforts, we fail to reproduce TENSORSCOPE's PyTorch fuzzer, so we do not report its API coverage, code coverage, and bug detection results on PyTorch.
Additionally, we use data from FreeFuzz\cite{wei2022free} and Docter\cite{xie2022docter} to support Orion's performance.

\noindent \textbf{API and Code Coverage.} 
Table \ref{api_and_code_cov} summarizes the API and code coverage achieved by all API-level techniques on both TensorFlow and PyTorch.
\sys demonstrates comprehensive coverage, supporting 3995 APIs in TensorFlow and 1868 APIs in PyTorch.
When compared to the best-performing baseline, Orion, \sys improves PyTorch's API coverage by 6.68\%, while maintaining comparable coverage for TensorFlow.
In terms of code coverage, \sys significantly outperforms the best baseline, TitanFuzz, with a 39.92\% improvement for TensorFlow and a substantial 98.20\% increase for PyTorch.
Since MoCo is a model-level fuzzer, we compare it only on the APIs it supports to ensure a fair comparison. 
As MoCo also supports Jittor, we include it in the comparison as well.
Table \ref{api_and_code_cov_moco} presents the comparison between MirrorFuzz and MoCo.
In terms of API and code coverage, MirrorFuzz outperforms MoCo in both aspects.
These improvements come from our carefully designed prompt templates, which use shared bugs and few-shot CoT to guide the LLM in generating more diverse and realistic code snippets.
This facilitates complex API interactions, covering essential processes such as model building, training, and serialization, enabling deeper framework exploration.
\begin{table}[t]
\caption{ Comparison of bug detection time efficiency on 400 APIs among Orion, TitanFuzz, and MirrorFuzz.}
\label{table:time_efficiency}
\resizebox{\columnwidth}{!}{
\begin{tabular}{@{}l|ccc|ccc@{}}
\toprule
\multicolumn{1}{c|}{\multirow{2}{*}{}} & \multicolumn{3}{c|}{\textbf{TensorFlow}} & \multicolumn{3}{c}{\textbf{PyTorch}} \\ \cmidrule(l){2-7} 
\multicolumn{1}{c|}{} & \textbf{\begin{tabular}[c]{@{}c@{}}APIs \\ Tested\end{tabular}} & \textbf{\begin{tabular}[c]{@{}c@{}}Bug \\ Found\end{tabular}} & \textbf{\begin{tabular}[c]{@{}c@{}}Testing \\ Duration\end{tabular}} & \textbf{\begin{tabular}[c]{@{}c@{}}APIs \\ Tested\end{tabular}} & \textbf{\begin{tabular}[c]{@{}c@{}}Bug \\ Found\end{tabular}} & \textbf{\begin{tabular}[c]{@{}c@{}}Testing \\ Duration\end{tabular}} \\ \midrule
\textbf{Orion} & 148/400 & 14 & 5h & 172/400 & 6 & 5h \\ \midrule
\textbf{TitanFuzz} & 287/400 & 2 & 5h & 231/400 & 2 & 5h \\ \midrule
\textbf{MirrorFuzz} & 385/400 & 27 & 5h & 400/400 & 11 & 4h23min44s \\ \bottomrule
\end{tabular}
}
\end{table}

\begin{figure}[t]
    \centering
    \includegraphics[width=1\linewidth]{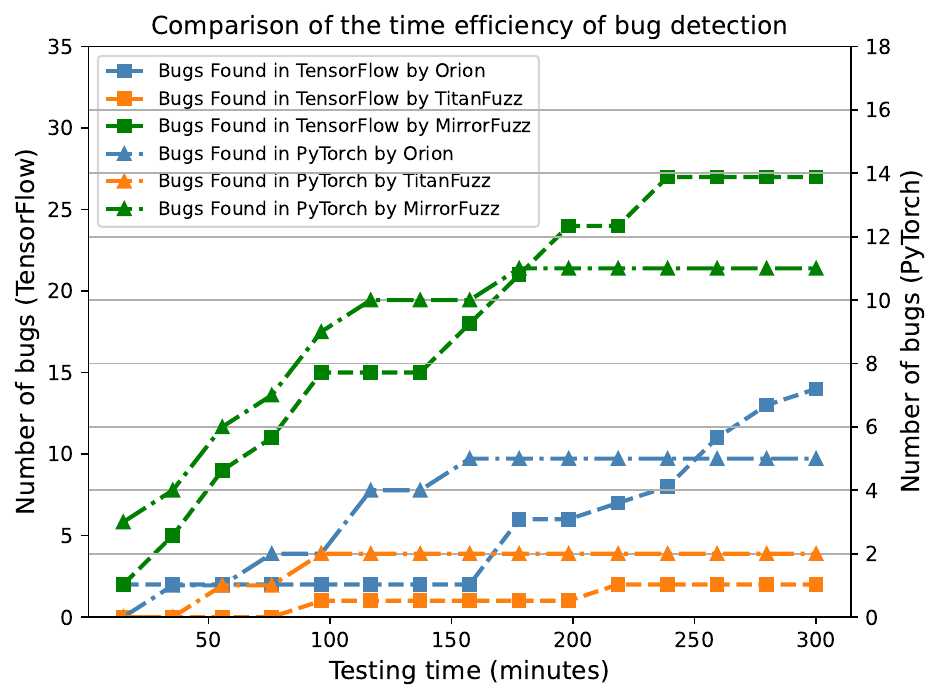}
    \caption{Comparison of bug detection time efficiency on 400 APIs.}
    \label{fig:bug_detection_time_efficiency}
\end{figure}

\begin{figure*}
    \centering
    \subfloat[PyTorch  1.31.1]{\includegraphics[width=0.25\textwidth]{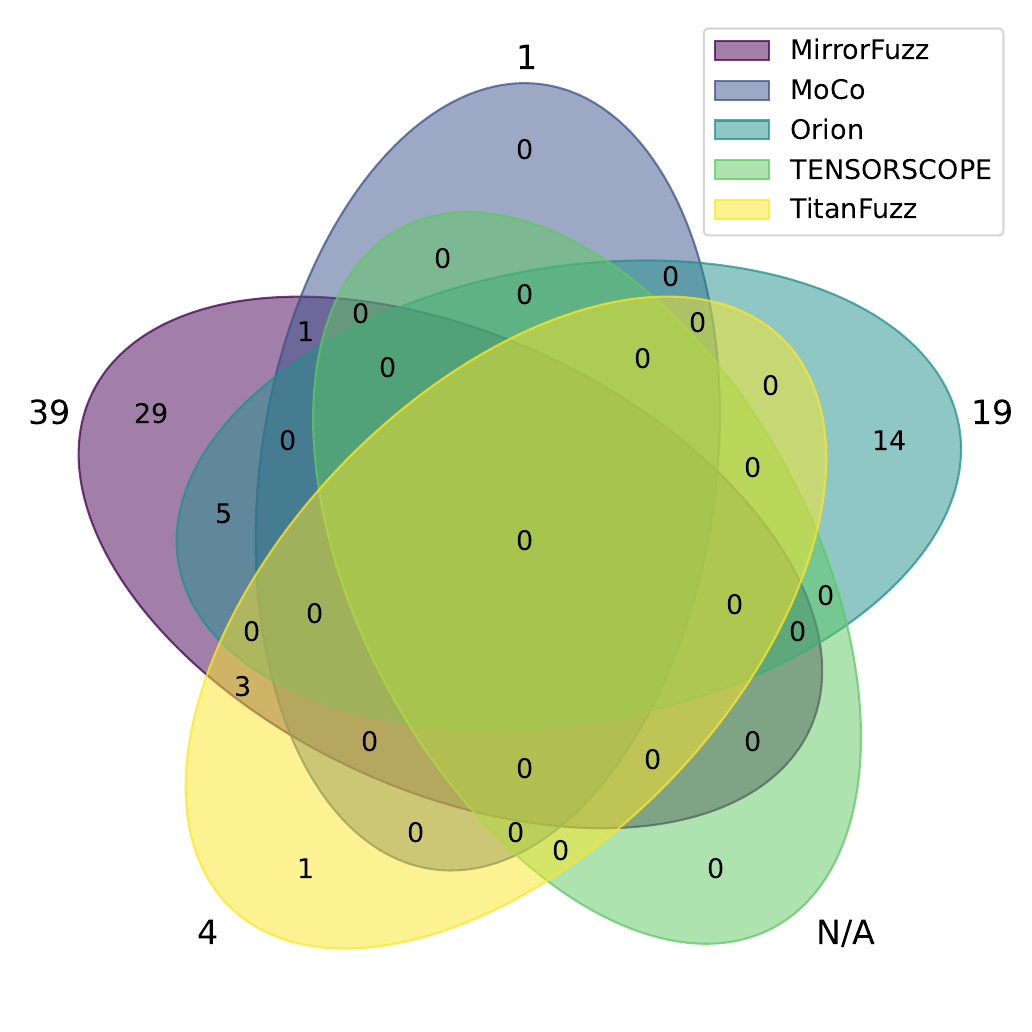}} 
    \subfloat[PyTorch  2.4.1]{\includegraphics[width=0.25\textwidth]{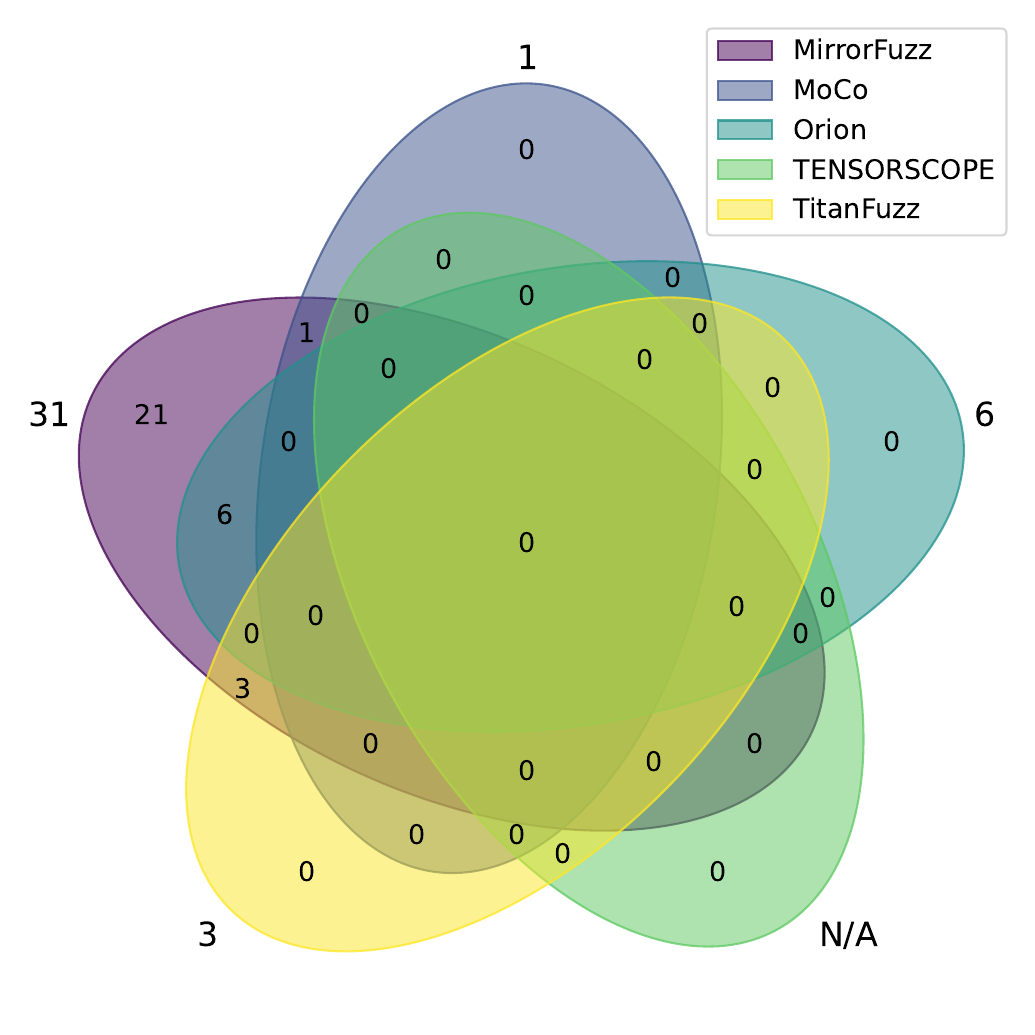}} 
    \subfloat[TensorFlow  2.10.0]{\includegraphics[width=0.25\textwidth]{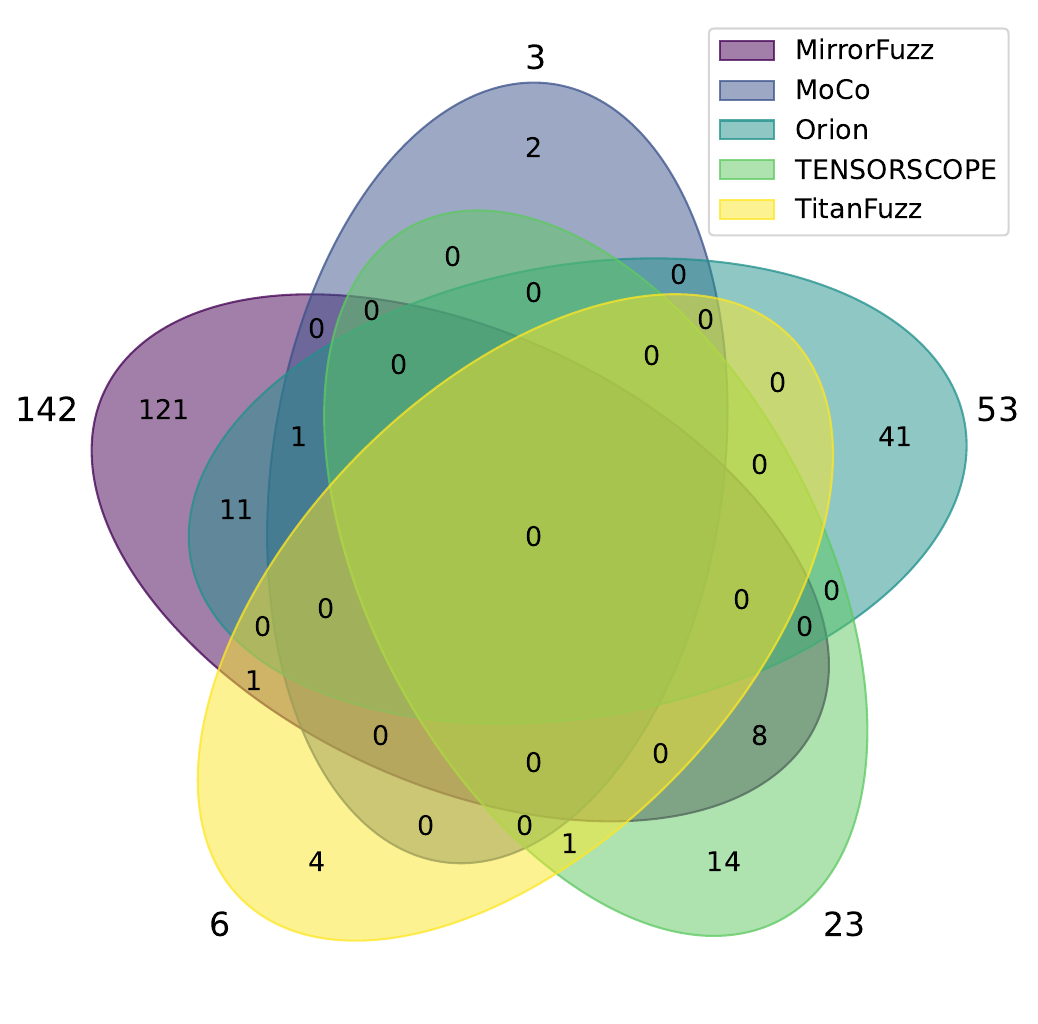}} 
    \subfloat[TensorFlow  2.17.0]{\includegraphics[width=0.25\textwidth]{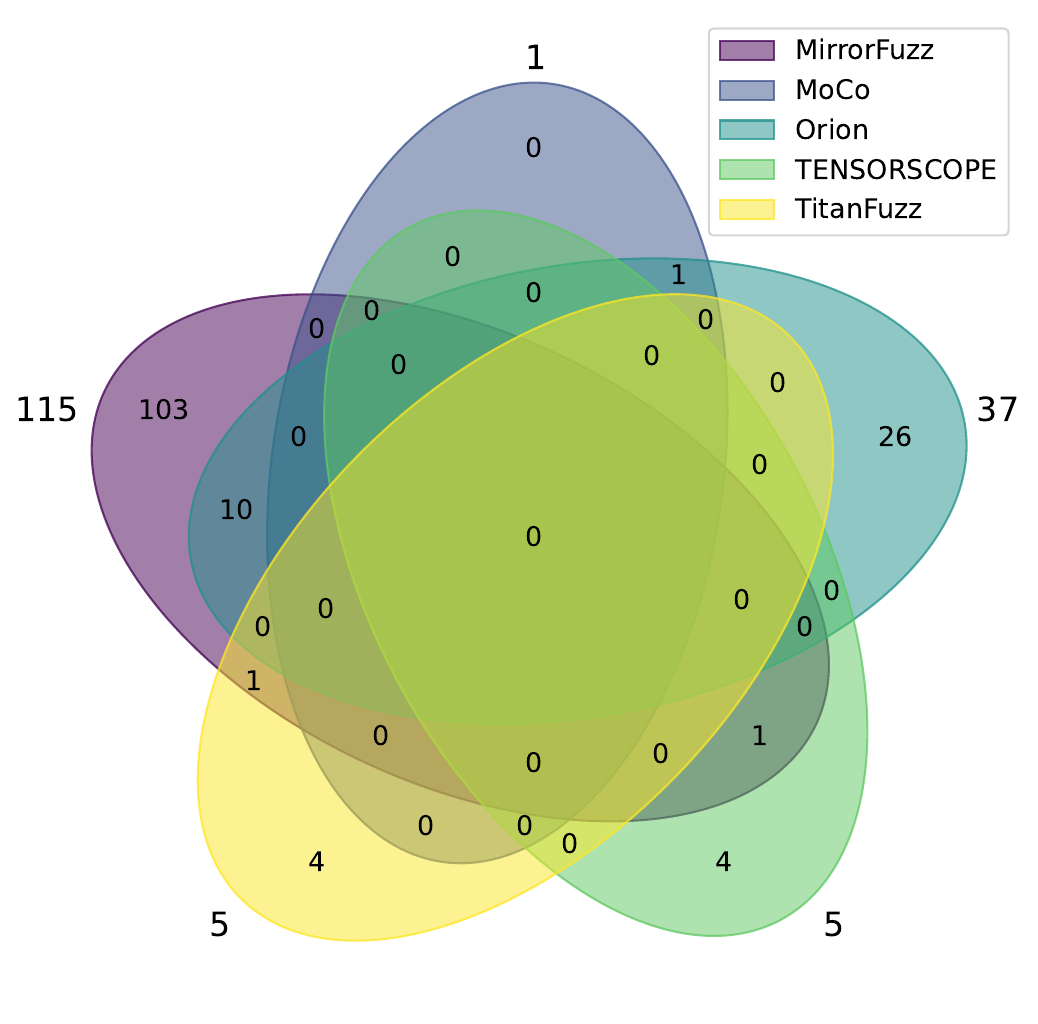}} 
    \caption{Venn diagram of unique crashes on TensorFlow and PyTorch releases.}
    \label{fig:comparison_bug}
\end{figure*}
\chadded[id=osw]{
\noindent \textbf{Bug Detection.} 
We compare the crash detection time efficiency of MirrorFuzz against baseline methods. Note that we exclude inconsistency bugs, as crashes are easier to measure and serve as a good approximation of bug-finding capabilities \cite{deng2023large}. Following the approach of TensorScope, we randomly select 400 APIs each from TensorFlow and PyTorch. We then ran each baseline tool with its default settings for a fixed duration of 5 hours. Since TensorScope only supports low-level APIs in TensorFlow, and MoCo supports a limited number of APIs, they are excluded from the efficiency comparison.
As shown in Table \ref{table:time_efficiency}, within the same time budget, MirrorFuzz tests 385 APIs in TensorFlow and completes testing all 400 APIs in PyTorch, outperforming other baselines.
Figure \ref{fig:bug_detection_time_efficiency} further illustrates the dynamics of bug discovery of the different tools over time. MirrorFuzz also reports the most unique bugs, finding 27 in TensorFlow and 11 in PyTorch. 
This demonstrates the advantage of MirrorFuzz in terms of bug detection efficiency.}

Furthermore, We compare the crash detection performance of MirrorFuzz and the baselines across different versions of TensorFlow and PyTorch.
Specifically, we conduct bug detection with each baseline’s default settings and its respective default time budget.
Figure \ref{fig:comparison_bug} shows a comparison of \sys and state-of-the-art fuzzers in terms of the number of bugs detected across different versions of TensorFlow and PyTorch.
\sys detects 39 and 31 bugs in PyTorch versions 1.13.1 and 2.4.1, respectively, and identifies 142 and 115 bugs in TensorFlow versions 2.10.0 and 2.17.0.
Compared to the baselines, \sys demonstrates superior performance by uncovering the highest number of crashes in both the earlier and current versions of TensorFlow and PyTorch.
Remarkably, despite the extensive prior testing of these frameworks, \sys uncovers a significant number of new bugs.
This success is attributed to the shared bug phenomenon discovered in our work, enabling \sys to synthesize code that triggers similar bugs across analogous APIs within and between frameworks, thereby revealing previously undetected flaws.

\section{Threats to Validity}
\label{section:Discussion}

\noindent \textbf{Internal.}
In our approach, we employ embedding models to extract semantic features from API metadata, and leverage LLMs for buggy API recognition and test case generation.
Differences between frameworks can introduce semantic ambiguity, which may limit the effectiveness of embedding models. 
To address this issue, we simultaneously consider text and semantic similarity and design an optimization algorithm that combines different thresholds to mitigate this effect.
Given the inherent unpredictability of LLMs, we strictly preprocess the crawled data and carefully design few-shot CoT prompt templates to enhance their accuracy and reliability in two distinct tasks.
Another internal threat is the construction of the ground-truth. To ensure its reliability and accuracy, we assign three authors to handle the process for each build and conduct cross-validation.

\noindent \textbf{External.}
The threats to the external validity of MirrorFuzz come from the tested DL frameworks. 
We select four commonly used DL frameworks, i.e., TensorFlow, PyTorch, OneFlow, and Jittor, as our research subjects.
Moreover, our fuzzing approach is generalizable to other frameworks.
The structure of documentation pages varies across different frameworks.
Users only need to adapt the provided crawler code to match the target documentation format in order to collect API metadata.
We provide a flexible GitHub issue crawler, and subsequent buggy API recognition, similar API matching, and test case synthesis are automatic and universal.

\section{Related Work}
To ensure the reliability of upper-layer applications built on DL frameworks, many studies\cite{cao2022understanding, chen2023toward,guan2023comprehensive,tambon2024silent} reveal and analyze bugs within these frameworks.
Furthermore, several fuzzing techniques are proposed to address these issues, aiming to enhance the security of the frameworks. These techniques are categorized into model-level fuzzing \cite{pham2019cradle,wang2020deep,guo2020audee,gu2022muffin,wang2022eagle,liu2023generation,liu2023nnsmith,liu2023neuri,go2024towards} and API-level fuzzing \cite{zhang2021duo,wei2022free,deng2022fuzzingDeepREL,xie2022docter,christou2023ivysyn,yang2023fuzzing,shi2023acetest,deng2023differential,deng2023large,deng2024largefuzzgpt,shirihistory,guan2024large}, depending on the type of test cases provided to the DL framework.

\noindent \textbf{Model-level Fuzzing.}
Model-level fuzzing detects bugs in the framework implementation by using a complete DL model as input.
These DL models consist of a range of APIs that cover various neural network functions.
CRADLE \cite{pham2019cradle} is a pioneering work in performing fuzzing across DL frameworks, using differential testing to detect and locate bugs by running pre-trained models across different DL backends supported by Keras \cite{keras}, such as TensorFlow \cite{abadi2016tensorflow}.
AUDEE\cite{guo2020audee} adopts search-based strategies and conducts cross-reference checks to detect behavioral inconsistencies across multiple DL frameworks. 
LEMON\cite{wang2020deep} and Muffin\cite{gu2022muffin} propose various mutation strategies that further augment CRADLE to generate diverse DL models. 
Later, these studies \cite{wang2022eagle, liu2023generation, liu2023nnsmith, liu2023neuri} propose different strategies to generate diverse DL models for bug detection in frameworks by differential testing.
Recently, DeepConstr\cite{go2024towards} refines API constraints by evaluating whether the complementary set of constraints generates valid test cases.
In contrast, \sys leverages LLM to synthesize test cases for APIs, enabling coverage of a broader range of APIs.
By referencing real-world buggy code snippets, \sys generates diverse API sequences, facilitating deeper testing of DL frameworks.

\noindent \textbf{API-level Fuzzing.}
Compared to model-level fuzzing, API-level fuzzing focuses on detecting bugs within individual APIs.
However, the complex constraints inherent in APIs pose a challenge for generating arbitrary test cases.
Consequently, these studies \cite{wei2022free,xie2022docter,deng2022fuzzingDeepREL, shi2023acetest} have made considerable efforts to generate valid or diverse test cases through relation reasoning or constraint extraction.
Then, IvySyn\cite{christou2023ivysyn} performs type-aware and mutation fuzz testing on low-level kernel code in DL frameworks to detect vulnerabilities.
TENSORSCOPE\cite{deng2023differential} identifies corresponding APIs by leveraging model transformation rules and performs cross-framework testing.
TitanFuzz\cite{deng2023large} leverages zero-shot prompts with LLMs to generate seeds and perform test case mutation.
FuzzGPT\cite{deng2024largefuzzgpt} uses historical bug data to prompt and fine-tune LLMs for generating unusual code.
Orion\cite{shirihistory} analyzes historical bugs to build fuzzing heuristic rules and generate corner-case test inputs for APIs.
Recently, YanHui\cite{guan2024large} proposes a domain knowledge-aware prompting method that uses LLMs to detect model optimization bugs.
\( D^3 \)\cite{wang2024d} uses distributed equivalence rules to test the distributed training and inference functionalities in DL frameworks.
Unlike the above studies, \sys reveals and analyzes the phenomenon of shared bugs between APIs across frameworks, leveraging this insight for early bug detection between frameworks.

\section{Future Work}
\label{section:FutureWork}

\chadded[id=osw]{In future work, we plan to extend our approach in several directions to address current limitations.}

\chadded[id=osw]{\noindent \textbf{Towards Adaptive Thresholds.}
In our work, we combine top-$k$ and a similarity threshold to filter candidate API matches.
The threshold is based on empirical observation, derived from evaluating candidate options and validating on API samples from multiple DL frameworks with varying functional complexities.
While this approach works in many cases, it may not be optimal in all scenarios.
In future work, we plan to explore the factors that influence the threshold \textit{H} and conduct a formal analysis of their relationships to construct an adaptive threshold that adapts to different scenarios.
In addition, we will leverage LLM-based agents to assist in filtering and refining candidate API matches, thereby improving the overall matching quality.}

\chadded[id=osw]{\noindent \textbf{Discovering Novel Bug Classes.}  
At present, our design relies on existing bug reports to detect similar bugs across DL frameworks.
This reliance may limit the system’s capacity to discover entirely novel bug classes (i.e., “unknown unknowns”).
To improve MirrorFuzz’s capability in finding new bugs, we adopted heuristic mutation operators and randomized scheduling to increase the diversity of generated test cases.
Furthermore, newly detected bugs are automatically incorporated into the bug records, allowing the scope of shared bug patterns to gradually expand over time.
In future work, we plan to explore adaptive mutation operators to further increase the diversity of test cases, thereby helping to uncover previously unknown bugs.
We also plan to explore vulnerability-aware fuzzing strategies for specific bug types, such as use-after-free, integer overflow, and buffer overflow, to further discover novel bug classes in DL frameworks.}

\chadded[id=osw]{\noindent \textbf{Improving Robustness and Adaptability with LLM Agents.}
The varying quality of historical bug reports poses a challenge for LLMs in the recognition of buggy APIs.
This problem is particularly common in DL frameworks with less mature community ecosystems, such as OneFlow and Jittor, where bug reports are often ambiguous or incomplete.
To mitigate this challenge, we employ targeted data preprocessing and carefully designed few-shot CoT prompts.
In future work, we plan to construct approaches driven by LLM-based agents, decomposing the workflow into sub-tasks such as bug report filtering, bug report normalization and enrichment, and final buggy API identification to enhance the analytical consistency and robustness of our approach.}

\chadded[id=osw]{
In addition, some components in the current design, such as keyword-based filtering, regex extraction, and Levenshtein Distance matching, were designed based on practical experience and have been shown to be effective in our experiments.
However, they may introduce brittleness and limit long-term extensibility.
With the rapid evolution of LLM capabilities, we plan to migrate the MirrorFuzz pipeline to a multi-agent-based paradigm.
In this paradigm, specialized agents replace these brittle components and handle key tasks such as buggy API recognition and similar API matching, to enhance the robustness and adaptability of MirrorFuzz.}

\section{Conclusion}
\label{section:Conclusion}
In this paper, we reveal that bugs found in the APIs of one framework may pose a threat to similar APIs within and across frameworks.
Based on this insight, we propose \sys, a novel approach that leverages shared bugs for early bug detection.
\sys employs LLM to identify buggy APIs from bug reports and synthesizes targeted test cases for similar APIs to induce similar bugs.
We evaluate \sys on four popular DL frameworks and discover 315 bugs, 262 of which are newly discovered, with 52 assigned CNVD IDs.
Moreover, \sys outperforms state-of-the-art fuzzers in bug detection, with code coverage improvements of 39.92\% on TensorFlow and 98.20\% on PyTorch, respectively.

\IEEEtriggeratref{67}
\bibliographystyle{IEEEtran}
\bibliography{references}


 




\vfill

\end{document}